\documentclass[11pt,english]{article}
\pdfoutput=1
\usepackage{jheppub}
\usepackage{lscape}
\usepackage{amsmath}
\usepackage{slashed}
\usepackage{mathtools}
\usepackage{cleveref}
\usepackage{tcolorbox}
\usepackage{bbold}
\usepackage{tikz}
\usepackage{pgfplots}

\usepackage{subcaption}
\usepackage{amssymb,tikz}
\usetikzlibrary{positioning}
\usetikzlibrary{shapes.geometric, arrows}

\def\ran{\text{Ran}\,}
\def\ker{\text{Ker}\,}
\def\tr{\text{tr}\,}

\DeclarePairedDelimiter{\norm}{\lVert}{\rVert}
\newcommand{\FigScaleFact}{0.29}
\newcommand{\FigSize}{1.7}

\begin{document}

\title{Non-equilibrium and equilibrium thermodynamic foundations of the 2D toric code within the SEAQT framework}

\author[a,c]{Cesar Damian, } 
\author[b,c]{Adriana Salda\~na-Robles, } 
\author[c]{Michael von Spakovsky}

\affiliation[a]{Departamento de Ingenier\'ia Mec\'anica, Universidad de Guanajuato, Carretera Salamanca-Valle de Santiago Km 3.5+1.8 Comunidad de Palo Blanco, Salamanca, Mexico.}
\affiliation[b]{Departamento de Ingenier\'ia Agr\'icola, Universidad de Guanajuato, Carretera Irapuato-Silao Km 9, Irapuato, Mexico.}
\affiliation[c]{Mechanical Engineering Department, Virginia Tech, Blacksburg, VA 24061, USA.}


\date{\today}

\abstract{The 2D toric code is a prototypical example that exhibits non-trivial topological properties and a ground state possessing a non-trivial topological order. Until now, all the cases studied in the literature have been in the stable equilibrium regime, leaving the relaxation towards this state unexplored. In the present work, the dynamics of the toric code towards stable equilibrium is explored within the steepest-entropy-ascent quantum thermodynamic (SEAQT) framework. Under these dynamics, information correlations such as the relative entropy, logarithmic negativity, magnetization, coherent information, and geometric entropy are described out of equilibrium, permitting a connection to be made between the out-of-equilibrium dynamics and these information measures.
}
\arxivnumber{}


\maketitle

\section{Introduction}

In the realm of quantum mechanics and information theory, stabilizer codes play a pivotal role. A quintessential example of such a stabilizer code is the 2D toric code, which is characterized by stabilizer operators that are elements of the Pauli group, encapsulating the simplest topological order denoted as $\mathbb{Z}_2$. This topological order is intrinsically linked to the von Neumann entropy, a metric that quantifies the entanglement of the ground state, which upon an appropriate choice of boundary is calculated as $\gamma  = \ln \mathcal{D}$ \cite{Kitaev:2006} where $\mathcal{D}$ stands for the total quantum dimension. For the 2D toric code, this dimension is 2. However, this topological order is susceptible to perturbations, which can induce a phase transition of the topological state \cite{Kitaev:2006,Kitaev:2002}.

The 2D toric code can be understood as the sum of two gauge theories such that the partition function factorizes in the two bases corresponding to these classical gauge theories, and as a consequence, there exist two temperatures at which excitations of each gauge theory proliferate \cite{Homeier:2021}.  It has been shown by employing entanglement negativity that the minimum temperature at which any of the two excitations proliferates tends to destroy the topological order \cite{Lu:2020}. Furthermore, the response of the 2D toric code to external perturbations has been extensively studied by checking several indicators of the topological order at stable equilibrium \cite{Fan:2023}. These indicators include the relative entropy, coherent information \cite{Schumacher:1996,Schumacher:2002,Horodecki:2007}, entanglement negativity \cite{Hart:2018,Lu:2023}, and phase changes due to magnetization \cite{Castelnovo:2008a}.

Delving deeper into the dynamics of quantum states, the steepest-entropy-ascent quantum thermodynamic (SEAQT) framework postulates that thermodynamic irreversibilities are inherently tied to the evolution of the state space \cite{Beretta:2009,beretta2010maximum,Beretta:2014,Li:2016a,li_steepest-entropy-ascent_2018,Li:2016b,  Beretta:2020}.  For instance, a flip-change (relaxation) is integral to the system dynamics whenever a heat interaction with an environment occurs whereas pure dephasing is tied exclusively to entropy generation internal to the system  \cite{Montanes:2022a,Montantez:2020a}. Among its many applications, the SEAQT framework has been effectively utilized to model and understand, for example, the atomistic spin relaxation processes in materials such as body-centered cubic iron \cite{Yamada:2019} as well as the relaxation of non-resonant excitation transfer processes \cite{Morishita:2023}, offering new insights into the far-from-equilibrium thermodynamic behavior of atomistic systems. The SEAQT framework has also been successfully applied to mesoscopic and macroscopic problems, aiding, for example, in the prediction of polymer folding behavior, the analysis of defects and stability in annealing kinetics, and understanding the behavior of lipid membranes \cite{Goswami:2021,Mcdonald:2023,Mcdonald:2023a}.

The robustness of the SEAQT framework bolsters confidence in its predictive capabilities. Its foundation on a deterministic dynamics oriented towards maximum entropy generation via the physical principle of steepest entropy ascent connects with non-equilibrium measurements, whose importance in particular has been noted in \cite{Jarzynski:1997,Jarzynski:1997a,Seifert:2012,li_steepest-entropy-ascent_2018}, and with fluctuation theorems in stochastic thermodynamics \cite{parrondo2015thermodynamics} that can be written in terms of the fluctuations of the non-equilibrium and equilibrium free energy. As a consequence, this physical principle, suggesting a unique, most probable evolutionary path towards stable equilibrium in any physical system, underscores the  formal application and theoretical underpinning of the SEAQT framework.

In the  present work, the dynamics towards stable equilibrium of the 2D toric code 2D is studied for various lattice sizes and information measures as well as for the code's connection to non-equilibrium and equilibrium thermodynamic properties. In Section \ref{sec:basics}, the standard notation of the toric code is presented as well as the perturbation methods employed in the manuscript to generate initial states.  Section \ref{sec:seaqt_ind}  presents a detailed derivation of the SEAQT equation of motion with emphasis on its fluctuation dissipation formulation and its prediction of stable equilibrium states. In Section \ref{sec:correlation}, the information measures employed are presented as well as their non-equilibrium evolution in the SEAQT framework. Section \ref{sec:examples} then provides four examples of lattices and the evolution towards stable equilibrium of the information measures in the SEAQT framework. This is followed in Section \ref{sec:thermo} with a thermodynamic explanation of the relaxation process towards stable equilibrium. Section \ref{sec:concl} concludes with a number of conclusions among which is the fact that a gradual decline in the topological properties of the toric code occurs when the irreversibilities dictated by the second law of thermodynamics are considered and the level of perturbation is increased. 

\section{\label{sec:basics}Basics of the 2D toric code}
The 2D toric code is composed of a lattice $\mathcal{L}_2$ that discretizes a two-dimensional torus together with a two-dimensional Hilbert space $\mathcal{H}_2$ constructed from the j-th edges of $\mathcal{L}_2$. Thus, the total Hilbert space is given by
\begin{equation}
    \mathcal{H} = \mathcal{H}_2 \otimes \ldots \mathcal{H}_2 \,,
\end{equation}
where each $\mathcal{H}_2$ is related to the physical qubits associated with each edge of the lattice. 

The Hamiltonian operator is defined as
\begin{equation}
    \hat H = - \sum_v \hat A_{s_v} - \sum_f  \hat B_{p_f} ,
\end{equation}
where the Hermitian star operator $\hat A_{s_v}$ and plaquette operator $\hat B_{p_f}$ commute with the Hamiltonian and are defined as
\begin{equation}
    \hat A_{s_v} = \prod_{j \in S_s} \hat \sigma^x_j \,, \quad \hat B_{p_f} = \prod_{j \in S_p} \hat \sigma_j^z
\end{equation}
Here, $\hat \sigma_j^x$ and $\hat \sigma_j^z$ are the $x-$ and $z-$Pauli matrices and
\begin{equation}
    \hat \sigma_j^{x,y,z} = I_2 \otimes \ldots \otimes I_2 \otimes \hat \sigma^{x,y,z} \otimes I_2 \otimes \ldots \otimes I_2 \,.
\end{equation}

This code defines a stabilizer code where the stabilizer operators $\hat A_{s_v}$ and $\hat B_{p_f}$ define the protected space such that
\begin{equation}
    \mathcal{L} = \{ |\xi \rangle \in \mathcal{H} :  \hat A_{s_v} |\xi \rangle = \hat B_{p_f} |\xi \rangle = |\xi \rangle, \forall s,p \}
\end{equation}
However, since $\hat A_{s_v}$ and $\hat B_{p_f}$ share two edges, they always commute, and the periodic boundary conditions ensure that
\begin{equation}
    \prod_{\forall v} \hat A_{s_v} =\prod_{\forall f}  \hat B_{p_f} = \mathbb{1}
\end{equation}
Thus, not all of the $\hat A_{s_v}$ and $\hat B_{p_f}$ are independent so that it follows that the dimensionality of the protected space is $\text{dim}\, \mathcal{L} = 2^{2}$. In addition, since the ground states coincide with the protected space, the ground space is four-fold degenerate. 

Now, defining the group generated by the independent $\hat A_{s_v}$ operators as $\mathcal{A}$ ($|\mathcal{A}| = 2^{k^2-1}$) where $k$ is the number of edges in the lattice representing the geometry of the 2D torus, the elements $\hat g \in \mathcal{A}$ become products of closed contractible strings. Furthermore, the $\omega_i$ for $i=1,2$ are defined as the loops of non-contractible paths of the torus so that the closure $\bar N$ is the group generated by $\mathcal{A}$ and $\omega_i$. As a result, if $S$ is an orthonormal basis of $\mathcal{H}$ such that
\begin{equation}
  \mathcal{S} = \{ |\psi \rangle = | s_1 = 0, \ldots , s_{2^{k^2}} \rangle : s_j = 0, 1 \}  
\end{equation}
the ground state $|00 \rangle = | s_{1}, \ldots , s_{2^{k^2}} = 0 \rangle$. The corresponding ground state density matrix is then that of a maximally mixed state that can be constructed as
\begin{equation}
    \hat \rho_0 = \frac{1}{4} \prod_v \frac{\hat I+\hat A_{s_v}}{2} \prod_f \frac{\hat I+\hat B_{p_f}}{2} \,,
\end{equation}
This can be understood as the sum of all the loops in the lattice. In loop notation, this ground state can be written as
\begin{equation}
    \hat \rho_0 = \frac{1}{|\mathcal{A}|} \sum_{\hat{g},\hat{g}'} \hat{g} |00 \rangle \langle 00 | \hat{g}'
\end{equation}
where the sum is over all possible loops on the torus and $\hat{g}$ and $\hat{g}'$ are different loop operators made up of the direct product of $\hat{A}_{s_v}$ or $\hat{B}_{p_f}$ operators. Mathematically, this implies that the zero cohomology group furnishes the ground state.  Utilizing the replica trick \cite{calabreseEntanglemententropyquantum2004,prihadiReplicaTrickCalculation2023}, the geometric entropy of the ground state can be epressed as $S_0 = - \text{tr} \hat{\rho}_0 \ln \hat{\rho}_0 =  \ln | \mathcal A |$. As is shown below, the relaxation towards stable equilibrium leads to a geometric entropy which matches that expected at equilibrium.

Now, consider a bi-partition of the Hilbert space into $\mathcal{H}_A \otimes \mathcal{H}_B$. The reduced density matrix for $\mathcal{H}_A$ is then \cite{Hamma:2005gr}
\begin{equation} \label{eq:reducedrhoA}
\hat \rho_A = \frac{1}{|\mathcal{A}/\mathcal{A}_B|} \sum_{\hat g \in \mathcal{A}/\mathcal{A}_B , \hat g_A \in \mathcal{A}_A} \hat x_A | 00\rangle_A \langle 00 |_A \hat x_A \hat g_A  \,
\end{equation}
where we consider $\mathcal{A}_A = \{\hat g \in \mathcal{A} |\hat g = \hat g_A \otimes \hat{\mathbb{1}}_B \}$, $\mathcal{A}_B = \{\hat g\in \mathcal{A} |\hat g = \hat{\mathbb{1}}_A \otimes \hat g_B \}$ and the elements $\hat g \in \bar N$ are decomposed as $\hat g = \hat x_A \otimes \hat x_B$, where $\hat x_A$, $\hat x_B$   are the elements of the subgroups $\mathcal{A}_A$, $\mathcal{A}_B$, which act non-trivially only on subsystems $A$ and $B$ \cite{Hamma:2005gr,Kitaev:2005dm}. 

Finally, errors due to single spin flip are also considered here such that
\begin{equation} \label{eq:pert}
\tilde \rho = (1-p_x) \hat \rho_0 +p_x \hat \sigma_i^x \hat \rho_0 \hat \sigma_i^x \,,
\end{equation}
where $\hat \sigma_i^x$ is the Pauli operator acting on the $i$-th lattice, creating $m$ anyons on the plaquettes, and $p_x$ is the corresponding error rate. Phase errors are taken into account 
naturally in the SEAQT framework via the irreversible dynamics.

\section{\label{sec:seaqt_ind}SEAQT formulation for indivisible systems}
In the SEAQT framework,  irreversibilities are intrinsic to the system and lead via the SEAQT dynamics to changes in the density matrix in the steepest-entropy-ascent direction. The SEAQT equation of motion for an isolated simple quantum system is written as
\begin{equation} \label{eq:EOM}
    \frac{d\hat \rho}{dt} = -\frac{i}{\hbar} \left[ \hat H, \hat \rho  \right] - \hat D
\end{equation}
where the commutator of $\hat H $ and $\hat \rho$ is the so-called von Neumann term, while the second term on the right is the dissipative term, which is exprssed as
\begin{equation}
    \hat D = \left(\sqrt{\hat \rho} \hat E_{D} + \hat E_{D}^\dagger\sqrt{\hat \rho}\right) 
\end{equation}
For the system considered here, there are only two generators of the motion: the identity operator, $\hat I$, and the Hamiltonian, $\hat H$. As a result, 
\begin{equation} \label{eq:ed}
\hat E_{D} = \frac{1}{2 \tau}    \frac{ \left|  \begin{matrix} 
  \sqrt{\hat \rho} \hat B \ln \hat \rho      & \sqrt{\hat \rho}   &   \sqrt{\hat \rho} \hat H  \\
      (  \sqrt{\hat \rho} \hat B \ln \hat \rho |\sqrt{\hat \rho} )  & ( \sqrt{\hat \rho} | \sqrt{\hat \rho}  )  & \left( \sqrt{ \hat \rho} \hat H |\sqrt{\hat \rho} \right)\\
      (  \sqrt{\hat \rho} \hat B \ln \hat \rho | \sqrt{ \hat \rho}\hat H  )  & ( \sqrt{\hat \rho} |  \sqrt{\hat \rho} \hat H  )  & ( \sqrt{\hat \rho} \hat H | \sqrt{\hat \rho} \hat H  )\\
   \end{matrix}
   \right|}{\left|  \begin{matrix} 
      ( \sqrt{\hat \rho} | \sqrt{\hat \rho}  )  & ( \sqrt{ \hat \rho} \hat H |\sqrt{\hat \rho} )\\
       ( \sqrt{\hat \rho} |  \sqrt{\hat \rho} \hat H  )  & ( \sqrt{\hat \rho} \hat H | \sqrt{\hat \rho} \hat H  )\\
   \end{matrix}
   \right| } 
\end{equation}
where the determinant in the denominator is a Gram determinant that ensures the linear independence of the generators of the motion,  $( \hat F | \hat G ) = \frac{1}{2} \text{tr}\, ( \hat \rho  \{ \hat F, \hat G \})$ is the real scalar product, $\{ \cdot, \cdot \}$ is the anti-commutator given by $( \hat F^\dagger \hat G + \hat G^\dagger \hat F )$, and $\hat F$ and $\hat G$ are self-adjoint operators. Note that the idempotent operator $\hat B$ is equal to $\hat P_{\ran}$, which is the projector onto the range of $\hat \rho$ such that $\hat B^2 = \hat B$ and $\left[ \hat B, \hat  \rho \right] = 0$. The operator $\hat B$ ensures that the expectation value of the entropy, $\langle s \rangle$, is always well defined because the trace is restricted to non-zero eigenvalues of $\hat \rho$. This can be understood alternatively by noticing that $\hat B = \hat I - \hat P_\ker \hat \rho$ where $\hat P_{\ker \hat \rho}$ denotes the projector onto the null space of $\hat \rho$. As a result, $\hat B \ln \hat \rho = \ln \left( \hat \rho - P_{\ker \hat \rho}\right)$, which effectively projects out all the zero eigenvalues.

Now, computing the scalar products in Eq. (\ref{eq:ed}), $\hat E_D$ reduces to 
\begin{equation} \label{eq:ed1}
    {\hat E_{D}} = \frac{1}{2 \tau} \frac{
\left|
\begin{matrix}
\sqrt{\hat \rho} \hat B \ln \hat \rho  & \sqrt{\hat \rho} & \sqrt{\hat \rho} \hat H  \\
\text{tr} ( \hat \rho \hat B \ln \hat \rho ) & 1 & \text{tr} ( \hat \rho \hat H ) \\
\text{tr} ( \hat \rho (\hat B \ln \hat \rho) \hat H )  & \text{tr} ( \hat \rho \hat H )& \text{tr} ( \hat \rho \hat H^2 )
\end{matrix}
\right|
}{
\left|
\begin{matrix}
 1 & \text{tr} ( \hat \rho \hat H ) \\
 \text{tr} ( \hat \rho \hat H )& \text{tr} ( \hat \rho \hat H^2 )
\end{matrix}
\right|
}
\end{equation}
Using the expectation values
\begin{eqnarray}
\langle s \rangle &=& -\text{tr} ( \hat \rho \hat B \ln \hat \rho) = \text{tr} ( \hat \rho \hat S) \\  
\langle e \rangle &=& \text{tr} ( \hat \rho \hat H )  \\ \langle e s\rangle &=& -\text{tr} ( \hat \rho (\hat B \ln \hat \rho) \hat H ) = \text{tr} ( \hat \rho \hat S \hat H )\\
\langle e^2 \rangle &=& \text{tr} ( \hat \rho H^2 )
\end{eqnarray}
and performing standard row operations on the determinant in the numerator that include multiplying row 2 by $\sqrt{\hat \rho}$ and subtracting it from row 1 and multiplying row 2 by $\textrm{tr} (\hat \rho \hat H )$ and subtracting it from row 3 results in
\begin{equation} \label{eq:ed2}
    {\hat E_{D}} = \frac{1}{2 \tau} \frac{
\left|
\begin{matrix}
- \sqrt{\hat \rho} \Delta \hat S  & \hat 0 & \sqrt{\hat \rho} \Delta \hat H \\
- \langle s \rangle & 1 & \langle e \rangle \\
\langle s \rangle \langle e \rangle-\langle se \rangle & 0 & \langle e^2 \rangle - \langle e \rangle^2 \end{matrix}
\right|
}{\left|
\begin{matrix}
 1 & \langle e \rangle \\
 \langle e \rangle & \langle e^2 \rangle 
\end{matrix}
\right|}
\end{equation}
In this last expression, $\Delta \hat S = \hat S - \langle s \rangle \hat I$ and $\Delta \hat H = \hat H - \langle e \rangle \hat I$ are the fluctuations of the entropy and energy operators, while the quantities $A_{se} = \langle se \rangle - \langle s \rangle \langle e \rangle$ and $A_{ee} = \langle e^2 \rangle - \langle e \rangle^2$ are the entropy-energy and energy-energy fluctuations, respectively \footnote{In the following, the notation $A_{ij}$ is adopted for the covariance whenever $i\neq j$ and for the variance whenever $i=j$.  This fluctuation-dissipation form of the equations of motion was initially suggested by Beretta \cite{Beretta:1987,Beretta:2006,Beretta:2009,beretta2010maximum}.}. 

Finally, expanding the determinants in Eq. (\ref{eq:ed2}), the $\hat E_D$ operator is written as
\begin{equation} \label{eq:ed3}
{\hat E_{D}} = \frac{1}{2\tau }\left( \sqrt{\hat \rho} (\hat B \ln \hat \rho )  - \beta \langle f \rangle \sqrt{\hat \rho}+\beta \sqrt{\hat \rho}  \hat H  \right)   
\end{equation}
with
\begin{eqnarray}\label{eq:bf}
\beta &=& \frac{A_{se}}{A_{ee}} \\ \langle f \rangle &=&  \langle e \rangle -  \beta^{-1} \langle s \rangle
\end{eqnarray}
where $ \beta$ is the non-equilibrium thermodynamic beta, which at stable equilibrium is inversely proportional to the temperature, and $\langle f \rangle$ is the free energy, which is valid at both non-equilibrium and stable equilibrium \cite{PhysRevE.100.022141,parrondo2015thermodynamics}. The equation of motion, Eq. (\ref{eq:EOM}), can now be written as 
\begin{equation} \label{eq:seaqiso}
\frac{d \hat \rho}{dt}= -\frac{i}{\hbar}\left[ \hat H, \hat \rho \right] -\frac{1}{\tau} \left( \hat \rho ( \hat B \ln \hat \rho ) - \beta \hat \rho \langle f \rangle + \frac{1}{2}\beta
 \{\hat H, \hat \rho \} \right) 
\end{equation}
Defining a non-negative entropy operator as
\begin{equation}
    \hat s = \hat \rho \hat S,
\end{equation}
and the energy operator as
\begin{equation}
    \hat e = \frac{1}{2}
 \{\hat H, \hat \rho \} ,
\end{equation}
Eq. (\ref{eq:seaqiso}) can be expressed as 
\begin{equation}
    \frac{d \hat \rho}{dt} = -\frac{i}{\hbar} \left[ \hat H, \hat \rho  \right] - \frac{\beta}{\tau} \left(  \hat f - \hat \rho \langle f \rangle \right)
\end{equation}
where $\hat f = \hat e - \beta^{-1} \hat s$.  Thus, the time rate of change of the density matrix depends on two contributions: the first due to the symplectic dynamics compatible with the Schr\"odinger equation and the second due to the fluctuation of the non-equilibrium free energy, $\langle f \rangle$, resulting from the dissipation dynamics of the SEAQT framework.  

At stable equilibrium, Eq. (\ref{eq:seaqiso}) has as the unique  thermodynamic non-dissipative solution  of a Gibbs state given by
\begin{equation}
    \hat \rho = \frac{\hat B e^{-\beta \hat H} \hat B}{Z}
\end{equation}
where $\ln Z =-\beta^{-1} \langle f \rangle $ is the natural logarithm of the usual partition function, $\hat B = I$ is the projector operator onto the rank of $\hat 
\rho$, and $\beta$ is inversely proportional to the thermodynamic temperature. Note that when $\hat B$ is  different from the identity matrix  and $\left[ \hat B, \hat H \right] \neq 0$, the state is still non-dissipative but unstable, i.e., only locally as opposed to globally stable, according to Lyapunov \cite{Beretta:2009}.  

If interactions between the system and a  thermal reservoir are present \cite{Li:2016b,Holladay:2019}, an additional generator of the motion, the identity operator for the reservoir, is required, which expands the determinant in the numerator given in Eq. (\ref{eq:ed}) from a 3 $\times$ 3 to a 4 $\times$ 4 determinant. The Gram determinant in the denominator expands as well to a 3 $\times$ 3 determinant. Thus, the generators of the motion now are $\hat I$ for the system $S$, $\hat I_R$ for the reservoir $R$, and $\hat H_C = \hat H + \hat H_R$ for the composite system $C$. In this case, an equation of motion for reservoir $R$ is not needed since it is so large (i.e., consists of a very large number of degrees of freedom) that its state of stable equilibrium does not change as it interacts with system $S$. Thus, the only an equation of motion required is that for system $S$, which is written as
\begin{equation}\label{eq:seaqnoniso}
    \frac{d \hat \rho}{dt}= -\frac{i}{\hbar}\left[ \hat H, \hat \rho \right] - \frac{\beta_R}{\tau} \left( \hat f - \hat \rho \langle f \rangle\right) 
\end{equation}
where $\hat f = \hat e - \beta_R^{-1} \hat s$ and $\hat \rho$ and $\langle f \rangle = \langle e \rangle - \beta_R^{-1} \langle s \rangle$ represent the density operator and free energy of system $S$, respectively. 

Now, although the time rate of change of the energy of the composite system $C$, which is isolated, is zero, that is not the case for system $S$  since it is interacting with the reservoir. Multiplying Eq.~(\ref{eq:seaqnoniso}) by $\hat H$ and taking the trace, the symplectic term vanishes since $\text{tr} (\left[ \hat H, \hat \rho \right] \hat H )= 0$. Using the expressions given for $\hat f$, $\langle f \rangle$, $\hat e$, $\hat s$, $A_{ee}$, and $A_{se}$ given above, the rest of the equation reduces to  
\begin{equation}
  \frac{d\langle e \rangle}{dt}= \frac{A_{ee}}{\tau} \left(  \beta -  \beta_R \right)  
\end{equation}
where $\beta$ is given by Eq.~(\ref{eq:bf}) and $\beta_R = 1/k_BT_R$. Here, $T_R$ is the temperature of the reservoir and $k_B$ is Boltzmann's constant. As can be seen, as the system approaches mutual stable equilibrium with the reservoir, $\beta$ approaches $\beta_R$ so that, as expected, the rate of change in energy goes to zero at stable equilibrium. 

The time rate of change of the entropy can also be found from Eq. (\ref{eq:bf}) by multiplying this equation by $\hat S$ and taking the trace. Thus,
\begin{equation} \label{eq:dsdt}
    \frac{d\langle s \rangle}{dt} = - \frac{ \beta_R}{\tau} A_{fs} \,,
\end{equation}
where $A_{fs} = \langle fs \rangle - \langle f \rangle \langle s \rangle$ is the free energy-entropy fluctuation and $\langle fs \rangle = \langle se\rangle - \beta_R^{-1} \langle s^2 \rangle$. Note that the term $-\beta_R A_{f s}$ is strictly positive definite, which follows from the Gram determinant \cite{beretta2010maximum}. Furthermore, at stable equilibrium, both the $A_{fs}$ and $A_{ee}$ fluctuations are zero, which implies that the fluctuation components $\langle e^2 \rangle$ and $\langle f s \rangle$ factorize.

\begin{figure}
    \centering    \includegraphics[width=0.5\linewidth]{ 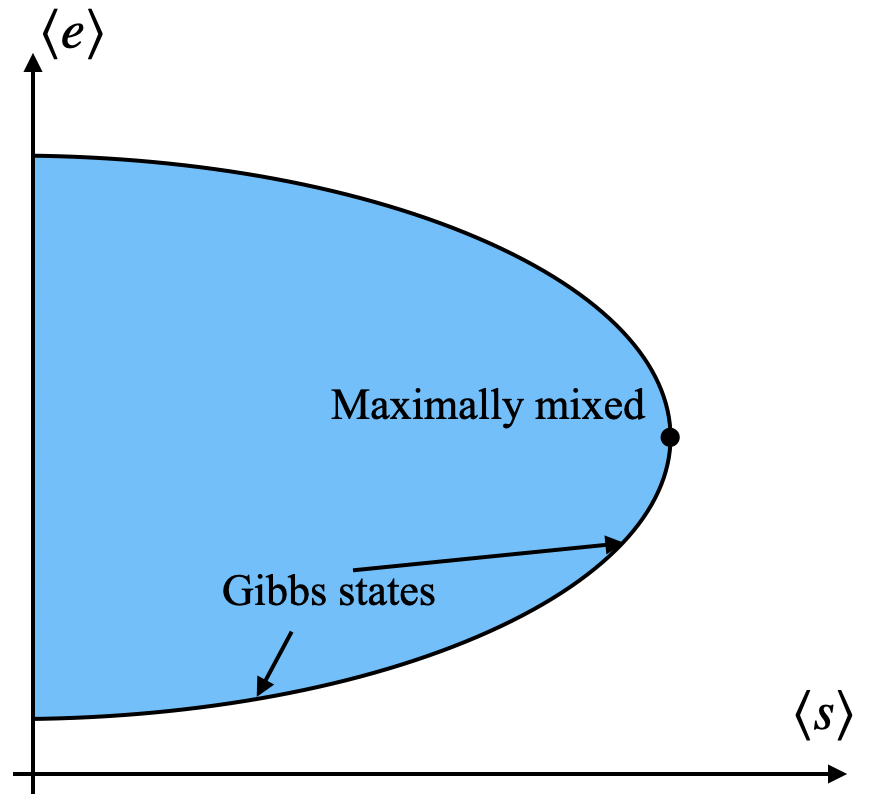}
    \caption{  \label{fig:esd}Energy-entropy diagram for representing the maximally mixed state (red circle) and all Gibbs states (blue curve).}
\end{figure}

Now, consider the case of high temperature, i.e., when $\beta_R$ approaches zero. In the limit, the product
\begin{equation}
   \lim_{ \beta_R \rightarrow 0} \beta_R \langle f \rangle = - \langle s \rangle 
\end{equation}
is well defined even when $\beta =0$ and a decrease of free energy means an increase in the entropy. Thus, the maximally mixed state as well as all of the Gibbs states can be represented with an $\langle e \rangle$ - $\langle s \rangle$ diagram as shown in Figure \ref{fig:esd}.

In the following sections, quantities directly related to the conservation of the topological properties of the toric code are checked in the context of the SEAQT framework as is how these quantities change in time. As will be seen, all the properties are conserved as long as the corresponding fluctuations factorize, which, of course, only occurs at stable equilibrium. Thus, the Gibbs states and the maximally mixed state of the toric code conserve the properties of the code since they are stable equilibrium states. However, as is also shown, along any kinetic path to stable equilibrium, the information contained in the toric code decreases in time due to irreversibilities, leading to unavoidable permanent loss of information.\\

\section{\label{sec:correlation}Correlation measures}
\subsection{Relative entropy}
To test the robustness of the topological order against perturbations, five correlation measures are employed. The first is the relative entropy, which is defined as
 \begin{equation}\label{eq:rel}
\mathcal{D} ( \hat \rho || \hat \rho_0 ) = \text{tr}\, \hat \rho \ln \hat \rho - \text{tr}\, \hat \rho \ln \hat \rho_0  
 \end{equation}
 where for the toric code, $\hat{\rho}_0$ is the ground state density matrix, which is a maximally mixed state. Its logarithm is proportional to the identity matrix, and since it is maximally mixed, it follows that the second term of Eq. (\ref{eq:rel}) only varies in time due to variations in the density matrix $\hat \rho$. Thus, the relative entropy change in time is given by
\begin{equation}
\frac{ d \mathcal{D} ( \hat \rho || \hat \rho_0 )}{dt} = - \frac{d \langle s \rangle}{dt} - k \, \text{tr}\, \frac{d \hat {\rho} }{dt}
\end{equation}
where $k$ is some real number. Now, since $\text{tr}\, \hat{\rho} =1$ and in the SEAQT framework, the trace of the density matrix is conserved, it follows that the second term vanishes and the relative entropy only changes due to the contribution of the rate change of the entropy. Then, based on Eq. (\ref{eq:dsdt}) for the case when the system is isolated and not interacting with a thermal reservoir, the rate of change of the relative entropy can be written as
\begin{equation}
\frac{d\mathcal{D}}{dt}(\hat{\rho} \,||\, \hat{\rho}_0) = \frac{\beta}{\tau} A_{fs},
\end{equation}
Thus, the rate of change of the relative entropy is the negative of the rate of change of the von Neumann entropy as already noted elsewhere\cite{Witten:2020}. Note that this rate of change is negative definite, indicating a tendency to decrease over time the distance between the ground state and any excited state. It is furthermore important to note that the entropy rate exhibits plateaus corresponding to partially canonical (unstable equilibrium) states, which act as attractors. As a consequence, the difference between $\hat{\rho}$ and $\hat{\rho}_0$ can only be sustained only as long as the states remain in these unstable equilibrium states.

\subsection{Logarithmic entanglement negativity}
The second correlation measure is the logarithmic entanglement negativity, which is defined with the trace 1-norm of the partially transposed density matrix $\norm{\cdot}_1$ defined as the sum of the absolute value of the eigenvalues. Given a Hilbert space $\mathcal{H} = \mathcal{H}_{A_1} \otimes \mathcal{H}_{A_2} \otimes \mathcal{H}_{B}$, where the subsystems $A_1$ and $A_2$ are chosen arbitrarily and $B$ is the complement of subsystems $A_1$ and $A_2$. The logarithmic entanglement negativity is then written as
\begin{equation}
    \mathcal{E} = \ln ||\hat \rho_A^{T_2}|| \,,
\end{equation}
where $\|\hat{\mathcal{A}}\| = \text{tr} \left( \sqrt{\hat{\mathcal{A}}^{\dagger} \hat{\mathcal{A}}} \right)$. Notice that the partial transpose of the density matrix contains negative eigenvalues and is, thus, not a physical representation of the density matrix but instead a mathematical object that keeps track of the quantum correlations in a given system. In particular, this correlation measure is zero if all the eigenvalues are positive and different from zero otherwise, permitting the detection of subsystem separability, e.g., subsystem $A$ from subsystem $B$. This measure furthermore provides an upper bound on the distillable entanglement of a bipartite state and can be expressed as a sum of local and non-local terms such that $\mathcal{E} = E_{N,\text{local}} + E_{N,\text{topo}}$ where $E_{N,\text{local}} = \alpha_{d-1} L_A^{d-3}+ \alpha_{d-3}L_A^{d-3}+ \ldots$. The non-local contribution, $E_{N,\text{topo}}$, is, unfortunately, not expressible as a functional of the local curvature, $L_A^{d-3}$,  along the entangling boundary. This correlation measure is usually employed to diagnose the topological order since it is used to check for quantum correlations in mixed-density matrices where the von Neumann entropy is dominated by classical correlations \cite{Cramer:2006}. To check for quantum correlations, the total system is divided into two subsystems $A \cup B$, and then subsystem $A = A_1 \cup A_2$ where $A_1$ and $A_2$ represent non-contractible cycles of subsystem $A$. This is done to preserve the topological information of the code.

\subsection{Magnetization}
Now, the third correlation measure considered here is the magnetization, which is expressed as
\begin{align}
    m = \frac{1}{N} \sum_{i=1}^N \text{tr}\, (\hat \sigma_i^z \hat \rho) = \frac{1}{N} \sum_{i=1}^N m_i\end{align}
where $N$ is the number of spins and $\hat \sigma_i^z$ is the $z$-Pauli matrix for spin ``$i$''. The time variation of $m$ then depends on the time variation of the density matrix, $\hat \rho$, which in the SEAQT framework is found for each $m_i$ by multiplying Eq. (\ref{eq:seaqiso}) by $\hat \sigma_i^z$ and taking the trace with the result that
\begin{equation}
    \frac{dm_i}{dt} = -\frac{i}{\hbar} \text{tr}(\hat \sigma_i ^z\left[\hat H, \hat \rho \right]) - \frac{1}{\tau} \left( \text{tr} (\hat \sigma_i^z \hat \rho \ln \hat \rho )-  \beta \langle f \rangle \text{tr}(\hat \sigma_i^z \hat \rho) + \frac{1}{2} \beta \text{tr}( \{ \hat H, \hat \rho \} \hat \sigma_i^z )\right)
\end{equation}
Defining $\langle s \sigma^z_i \rangle = -\text{tr} \left( \hat \sigma_i^z \hat \rho \ln \hat \rho \right)$ and $\langle e \sigma^z_i \rangle = \frac{1}{2} \text{tr} \left( \{ \hat H , \hat \rho \} \sigma_i^z\right)$, the local magnetization rate of change can be written as
\begin{equation} \label{eq:dmidt}
    \frac{dm_i}{dt} = -\frac{i}{\hbar} \text{tr}(\sigma_i ^z\left[ \hat H, \hat \rho \right]) - \frac{\beta}{\tau} A_{f \sigma_z^i} \,,
\end{equation}
where the free energy-Pauli matrix fluctuation $A_{f \sigma_i^z} = A_{e\sigma_i^z} -  \beta^{-1}A_{s\sigma_i^z}$ with $A_{e\sigma_i^z} = \langle e \sigma_i^z \rangle - \langle e \rangle \langle \sigma_i^z \rangle$ and  $A_{s\sigma_i^z} = \langle s \sigma_i^z \rangle - \langle s \rangle \langle \sigma_i^z \rangle$. Eq. (\ref{eq:dmidt}) indicates that the local magnetization is oscillating due to the symplectic dynamics while simultaneously dissipating its amplitude due to internal irreversibilities. This behaviour, indicates that locally each spin can see the other spins as a fictitious environment due to the irreversibilities modeled in the SEAQT framework. The total magnetization is then given by of all the fluctuations such that
\begin{equation}
    \frac{d m}{d  t} = -\frac{i}{N \hbar} \sum_{i=1}^N  \text{tr}(\hat \sigma_i ^z\left[ \hat H, \hat \rho \right])- \frac{\beta}{N \tau} \sum_{i=1}^N A_{f \sigma_i^z} \,.
\end{equation}

Now, consider the case where $\hat \rho$ is the ground state $\hat \rho_0$. Since the ground state is defined as a simultaneous eigenstate of all the vertex and plaquette operators it commutes with the Hamiltonian so that
\begin{equation}
    \text{tr}(\hat \sigma_i ^z\left[ \hat H, \hat \rho_0 \right]) = 0 \\
\end{equation}
In this case, the total magnetization rate reduces to
\begin{equation}
    \frac{dm}{dt} = - \frac{ \beta}{N\tau} \sum_{i=1}^N   A_{f \sigma_i^z}
\end{equation}
showing that the rate is a linear function of the sum of the fluctuations between the total energy of the system and each single magnetization. Furthermore, as the temperature increases, the rate of change of the total magnetization increases provided the sum is positive, i.e., that on average, variations in the energy and magnetization tend to increase or decrease together. However, if the sum is negative, then on average as the energy increases, the magnetization decreases. In this case, the rate of change of the total magnetization decreases with increasing temperature. Thus, in general, the total magnetization rate describes how the relative orientation among all the spins in the lattice change over time.  Within the SEAQT framework, this rate evolves towards zero due to the intrinsic irreversibilities present in the system. This result is indeed a generalization of previous studies \cite{Montantez:2020a}, where it is shown that via the SEAQT framework,  pure dephasing is directly related to the entropy generation.

\subsection{Coherent information}
The fourth correlation measure employed is the coherent information \cite{Lloyd:1996at}, $I_c$, which provides a check for the integrity of quantum memory. It considers the state of some isolated system $I$ compared with that of the system coupled to some reservoir $R$. The SEAQT framework models this as decoherence due to both amplitude damping as well to pure dephasing \cite{Montantez:2020a,Montanes:2022a}. The coherent information takes the form
\begin{equation}
    I_c ( I\langle R ) = \langle s \rangle^{I} - \langle s \rangle^{R}
\end{equation}
Then using Eq. (\ref{eq:dsdt}) for the case of an isolated system as well as for the case of the system interacting with a reservoir, the rate of change of the coherent information is written as
\begin{equation}
   \frac{d I_c ( I\langle R )}{dt} = -\frac{1}{\tau} \left(  \beta A_{fs}^I -  \beta_R A_{fs}^R \right) 
\end{equation}
where the dissipation time and dynamics is the same for both the isolated and non-isolated systems. This expression can be further reduced to
\begin{equation}
   \frac{d I_c ( I\langle R )}{dt} = -\frac{1}{\tau} A_{es} \left( \beta -\beta_R \right) \,. 
\end{equation}
For the case when the reservoir temperature is hotter than that of the system so that  $ \beta_R <  \beta$, the rate of change of the coherent information is negative since $A_{es}$ is positive definite. In this case, the rate of loss of coherent information decreases as $\beta$ approaches $\beta_R$, going to zero when the system comes to mutual stable equilibrium with the reservoir.  For the case when the reservoir temperature is colder than that of the system so that $ \beta_R > \beta$, the rate of change of the coherent information is positive in which case there is no loss of coherent information but in fact a gain, which decreases as $\beta$ approaches $\beta_R$.

\subsection{Geometric entropy}
The final correlation measure is the geometric entropy, which is defined as the von Neumann entropy of a reduced density matrix, $\hat \rho_J$. It is used to measure the correlations between the degrees of freedom of subsystem $J$ and the other subsystems designated here by $\bar J$. In the SEAQT framework, the equation of motion for a general quantum system of divisible systems is given by \cite{beretta2010maximum}
\begin{equation} \label{eq:eomgen}
    \frac{d \hat \rho}{dt} = -\frac{i}{\hbar} \left[ \hat H, \hat \rho \right] + \sum_{J=1}^M\left( \frac{D\hat \rho }{Dt} \right)^J\otimes \hat \rho_{\bar{J}}
\end{equation}
Consistent with Eq. (\ref{eq:eomgen}), the dynamics of the reduced density matrix, $\hat \rho_J=\textrm{tr}_{\bar J}\left(\hat \rho \right)$, is written as
\begin{equation}\label{eq:seaqto}
    \frac{d \hat \rho_J}{dt} = -\frac{i}{\hbar} \left[ \hat H_J, \hat \rho_J \right] - \frac{i}{\hbar} \text{tr}_{\bar J} \left( \left[ \hat V, \hat \rho \right] \right) + \left( \frac{D\hat \rho }{Dt} \right)^J
\end{equation}
where the symplectic dynamics is complemented by a possible interaction term of the Hamiltonian $\hat V$ along with the local dissipative term given by 
\begin{equation} \label{eq:DrhoDtJ}
     \left( \frac{D\hat \rho }{Dt} \right)^J = \left( \sqrt{\hat \rho_J}E_{D_J} + E_{D_J}^\dagger\sqrt{\hat \rho_J}\right)
\end{equation}
$E_{D_J}$ is explicitly written in he same manner as Eq. (\ref{eq:ed}) above  such that 
\begin{equation} \label{eq:edJ}
\hat E_{D_J} = \frac{1}{2 \tau_J}    \frac{ \left|  \begin{matrix} 
  \sqrt{\hat \rho_J} \hat B_J \ln \hat \rho_J      & \sqrt{\hat \rho_J}   &   \sqrt{\hat \rho_J} \hat H_J  \\
      (  \sqrt{\hat \rho_J} \hat B_J \ln \hat \rho_J |\sqrt{\hat \rho_J} )  & ( \sqrt{\hat \rho_J} | \sqrt{\hat \rho_J}  )  & \left( \sqrt{ \hat \rho_J} \hat H_J |\sqrt{\hat \rho_J} \right)\\
      (  \sqrt{\hat \rho_J} \hat B_J \ln \hat \rho_J | \sqrt{ \hat \rho_J}\hat H_J  )  & ( \sqrt{\hat \rho_J} |  \sqrt{\hat \rho_J} \hat H_J  )  & ( \sqrt{\hat \rho_J} \hat H_J | \sqrt{\hat \rho_J} \hat H_J  )\\
   \end{matrix}
   \right|}{\left|  \begin{matrix} 
      ( \sqrt{\hat \rho_J} | \sqrt{\hat \rho_J}  )  & ( \sqrt{ \hat \rho_J} \hat H_J |\sqrt{\hat \rho_J} )\\
       ( \sqrt{\hat \rho_J} |  \sqrt{\hat \rho_J} \hat H_J  )  & ( \sqrt{\hat \rho_J} \hat H_J | \sqrt{\hat \rho_J} \hat H_J  )\\
   \end{matrix}
   \right| } 
\end{equation}
Now, using a development similar to that used to arrive at Eq. (\ref{eq:ed3}) above, Eq. (\ref{eq:edJ}) can be written conveniently as
\begin{equation} \label{eq:edJ1}
  E_{D_J} = \frac{1}{2\tau_J }\left( \sqrt{\hat \rho_J}  \hat B_J \ln \hat \rho_J  - \beta_J \langle f \rangle_J \sqrt{\hat \rho_J}+ \beta_J \sqrt{\hat \rho_J} \hat H_J \right)  
\end{equation}
where 
\begin{equation}
     \beta_J = \frac{\langle es \rangle_J - \langle e \rangle_J \langle s \rangle_J}{\langle e^2 \rangle_J - \langle e \rangle_J^2} 
\end{equation}
\begin{equation}
    \langle f \rangle_J = \ \langle e \rangle_J -   (\beta_J)^{-1} \langle s \rangle_J
\end{equation}
and $\langle f \rangle^J$ is the local non-equilibrium free energy. Substitution of Eq. (\ref{eq:edJ1}) into Eq. (\ref{eq:DrhoDtJ}) then yields 
\begin{equation}\label{eq:seaqtsim}
 \left( \frac{D\hat \rho }{Dt} \right)^J = \frac{1}{\tau_J} \left( \hat \rho_J  \hat B_J \ln \hat \rho_J  -  \beta_J \hat \rho_J \langle f \rangle_J + \frac{1}{2} \beta_J 
 \{ \hat \rho_J , \hat H_J \} \right) 
\end{equation} 
This last expression can be written in terms of a free energy fluctuation using the entropy and energy functionals, i.e.,
\begin{equation}
\hat s_J = -\hat \rho_J \hat B_J \ln \hat \rho_J  
\end{equation}
and
\begin{equation}
    \hat e_J = \frac{1}{2} \{ \hat \rho_J, H_J \}
\end{equation}
as well as the free-energy operator $\hat f_J = \hat e_J - (\beta_J)^{-1} \hat s_J$. Thus, 
\begin{equation} \label{eq:DrhoDtJ1}
     \left( \frac{D\hat \rho }{Dt} \right)^J = \frac{\beta_J}{\tau_J} \left( \hat f_J- \hat \rho_J \langle f \rangle_J \right) 
\end{equation}
Eq. (\ref{eq:DrhoDtJ1}) indicates that the local non-dissipative dynamics of each subsystem $J$ evolves until the fluctuations of the local free energy go to zero. This is closely related to the hypoequlibrium idea presented by Li and von Spakovsky \cite{Li:2016b}.

Together Eqs. (\ref{eq:seaqto}) and (\ref{eq:DrhoDtJ1}) provide the equation of motion for the reduced density matrix in terms of the local perception of the free energy. Note that the trace term in  Eq. (\ref{eq:seaqto}) vanishes identically, since the unit trace of the reduced density matrix is conserved during the dynamics. For the 2D toric code, the ground state is four-fold degenerate and, as has been proven elsewhere \cite{Kitaev:2002}, each state has the same geometric (or entanglement) entropy. Thus, starting with the the ground state density matrix as the initial density matrix with entropy {\footnote{The geometric entropy associated with a pure state and a geometrical region  $J$ is the von Neumann entropy of its reduced density matrix \cite{Hamma:2005gr}.}}
\begin{equation} \label{eq:expectsJ}
    \langle s \rangle_J = - \text{tr}\, \left( \hat \rho_J\hat B_J \ln \hat \rho_J \right) \,
\end{equation}
the time rate of change of the entropy of $\langle s \rangle_J$ can be constructed from Eq. (\ref{eq:seaqto}). This is done by again noting that the trace term vanishes and by multiplying both sides of the equation by $\hat B_J\ln \hat\rho_J$ and taking the trace. Recognizing that since the density matrix of the reduced system $\hat \rho_J$ is still of trace one so that 
\begin{equation}
    \text{tr} \left(\frac{d}{dt} (\hat\rho_J \hat B_J\ln \hat \rho_J)\right) = \text{tr} \left(\hat B_J\ln \hat \rho_J \frac{d \hat \rho_J }{dt } \right) \; ,
\end{equation}
the time rate of change of the local entropy can be written as
\begin{equation} \label{eq:dsdt}
    \frac{d \langle s \rangle_J}{dt} = \text{tr} \, \left[ \left( \frac{D\hat \rho}{dt} \right)^J \hat B_J\ln \hat \rho_J \right]
\end{equation}
where the trace of the commutator term vanishes due to its antisymmetry. Substituting Eq. (\ref{eq:seaqtsim}) and using the definitions
\begin{eqnarray}
    \langle s^2 \rangle_J = \text{tr}_J  \left( \hat \rho_J (\hat B_J \ln \hat \rho_J) \ln \hat \rho_J \right) \,, \\
    \quad \langle e s \rangle_J = -\text{tr}_J \left( \hat \rho_J \hat H_J \ln \hat \rho_J \right)
\end{eqnarray}
the time rate of change of the entropy reduces to
\begin{equation}
    \frac{d\langle s \rangle_J}{dt} = -\frac{ \beta^J}{\tau^J} A_{fs}^J
\end{equation}
where $A_{fs}^J= \langle f s \rangle^J - \langle f \rangle^J \langle s\rangle^J$, and $\langle f s \rangle^J = \langle e s \rangle^J -  \beta^{-1} \langle s^2 \rangle^J$.  Thus, quite naturally, the SEAQT framework provides the rate of change of the geometric entropy for the case of divisible systems. 

Now, to compute the geometric entropy at stable equilibrium, the partially traced density matrix over subsystem $A$ of the ground state is computed through the replica trick \cite{calabreseEntanglemententropyquantum2004,prihadiReplicaTrickCalculation2023} such that
\begin{equation}
    \lim_{n\rightarrow 1}\frac{d}{dn} \text{tr}\hat \rho^n_A  = \text{tr} \hat \rho_A \ln \hat \rho_A \,,
\end{equation}
Then, raising Eq. (\ref{eq:reducedrhoA}) to the  $n^{th}$-power and recognizing that $(1/f)\sum_{\hat g'_A \in \mathcal{A}} \hat \rho_A \hat g'_A = (1/f) \hat \rho_Ad_A$, the reduced density matrix can be written as
\begin{equation}
\hat \rho_A^n = \left( \frac{d_A}{f} \right)^{n-1}  \hat \rho_A \,,    
\end{equation}
Taking the natural logarithm of both sides of this expression and the derivative with respect to $n$ results in
\begin{equation} \label{eq:SA}
    \hat \rho_A \ln \hat \rho_A = \ln \left( \frac{d_A}{f}\right) \hat \rho_A = -S_A \hat \rho_A \,,
\end{equation}
where $d_A = |\mathcal{A}_A|$, and $f= |\mathcal{A}|/|\mathcal{A}_B| = |\mathcal{A}|/d_B$ are geometric parameters of the toric code defined by the cardinality of $\mathcal{A}_A$ and $\mathcal{A}/\mathcal{A}_B$, respectively \cite{Hamma:2005gr}. For instance, if the partition $A$ takes into account only one spin, then $d_A = 1$ and $f=2$. 

To compute the time rate of change of $S_A$, the trace of the equation of motion, Eq. (\ref{eq:seaqto}), multiplied by $\ln \hat \rho_A$ is taken where the symplectic and interaction terms vanish for the reasons given previously. The result is
\begin{equation}
  \tau_A \frac{d S_A}{dt} = \text{tr}\, \left( \hat \rho_A \ln \hat \rho_A ( \hat B_A \ln \hat \rho_A )\right) -  \beta_A \left[ \langle f \rangle_A  S_A - \text{tr}\, \left( \hat \rho_A \hat B_A\ln \hat \rho_A H_A \right)\right]  
\end{equation}
Using Eqs. (\ref{eq:expectsJ}) and (\ref{eq:SA}) and the fact that $\langle e \rangle_J = \tr (\hat \rho_J \hat H_J)$, this last expression reduces to
\begin{equation}
 \tau_A \frac{d S_A}{dt} =\left[  - \beta_A \langle f \rangle_A   + \beta_A \langle e \rangle_A  - \langle s \rangle _A  \right] \ln \left( \frac{d_A}{f}\right) 
\end{equation}
Clearly, at stable equilibrium the time rate of change of $S_A$ is zero from which it follows that the time rate of change of the entropy, $\langle s \rangle_A$, of the ground state is zero. Thus the partial trace of the ground state is non-dissipative and stable, which implies that there is never a change in the boundary of the bi-partition (i.e., the boundary between $A$ and $B$), preserving its topological properties. \\

\section{\label{sec:examples}Explicit examples}
\subsection{Lattice of $\text{dim} \mathcal{H} = 2^2$}
\begin{figure}[htbp]
   \centering
 a)  \includegraphics[scale=\FigScaleFact]{ 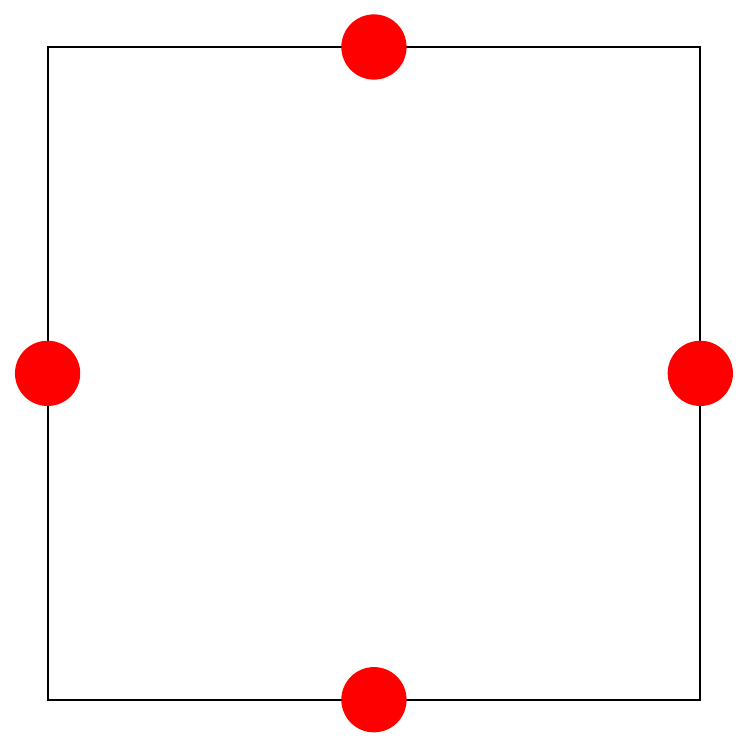} b)  \includegraphics[scale=\FigScaleFact]{ 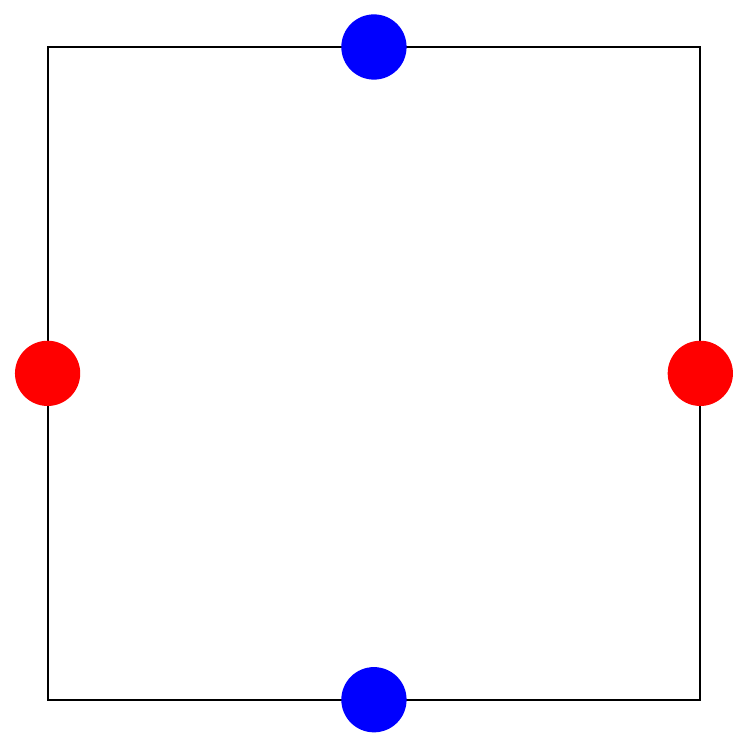} 
   \caption{Schematic of the lattice of a 2D toric for one plaquette in a) the ground state where the red circles represent the spin down particles in the computational basis, and in b) an excited state where the blue circles represent the spin up particles in the same basis.}
   \label{fig:lattice_1}
\end{figure}

Consider the simplest case of the 2D toric code, namely, that with $\text{dim} \mathcal{H} = 4$. For this case, the lattice takes the simple form shown in Fig. \ref{fig:lattice_1}.
Due to the periodic boundary conditions, there is only one star and one plaquette operator, and the ground state density matrix is simply
\begin{equation}
    \hat \rho_0 = \frac{1}{4} \left( \hat I + \hat A_1 \right) \left(\hat I + \hat B_1 \right)
\end{equation}
where $\hat A_1 = \hat \sigma_x^2 \otimes \hat \sigma_x^2 = I$ and $\hat B_1 = \hat \sigma_z^2 \otimes \hat \sigma_z^2$. This ground state is proportional to the identity operator and, thus, implies a maximally mixed state with rank operator equal to zero. This means that it is stable, i.e., non-dissipative. 

\begin{figure}[htbp]
   \centering
   \includegraphics[scale=\FigScaleFact]{ 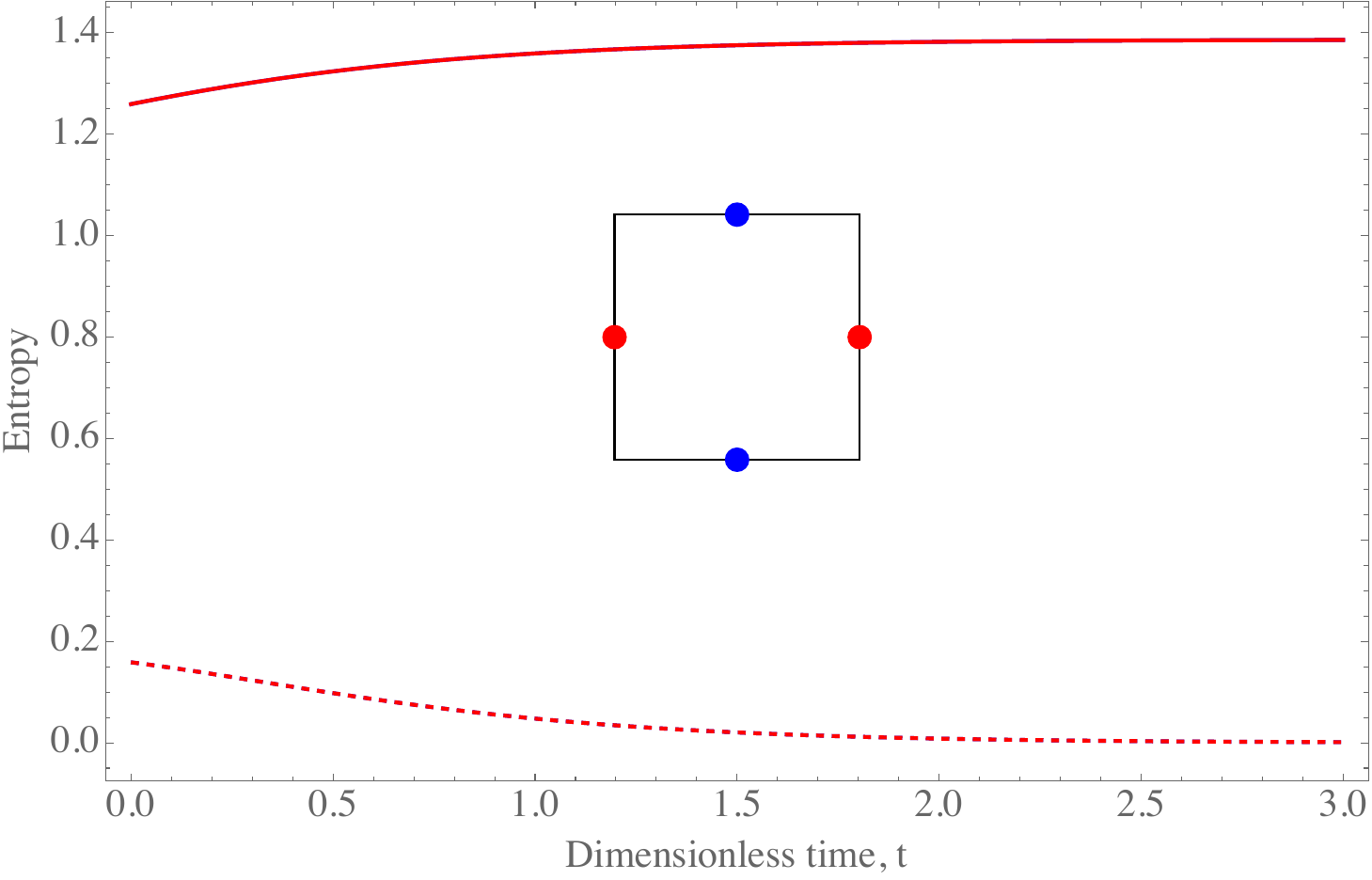} 
   \caption{ \label{fig:ent1} Entropy (solid curve) and rate of entropy change (dashed curve) of the 2D toric code with one plaquette as a function of dimensionless time where the blue curve represents the isolated system and the red curve the system interacting with a thermal reservoir; the difference in their evolutions is negligible. The initial state used is that based on $p_x = 0.4$.}
\end{figure}

For this simplest case, the symplectic dynamics vanishes since the density operator of the ground state is proportional to the identity operator as well as the Hamiltion operator. Thus, the density matrix commutes with the Hamiltonian operator. To check relaxation towards stable equilibrium, an initial excited (see Fig. \ref{fig:lattice_1}b)) state for the equation of motion is generated by first perturbing the identity operator using the Pauli operator $\hat \sigma_1^x$ after which the resulting state is perturbed once more in the direction of the entropy and energy of the orginal unperturbed state (see \cite{Montanez:2022b} for details of this perturbation method). This initial state based on $p_x = 0.4$ then evolves based on Eq. (\ref{eq:seaqiso}) for the isolated system and on Eq. (\ref{eq:seaqnoniso}) for the system interacting with a thermal reservoir. The evolutions of the entropy and rate of entropy change are shown in Fig. \ref{fig:ent1} for the case of the isolated system (blue curves) and for the case of the system interacting with a thermal reservoir (red curves). As seen in the figure, the red and blue curves lie almost directly on top of each other since the difference in the evolutions for these two cases is negligible. This is due to the fact that the symplectic dynamics vanishes and the intensive property $\beta$ in Eq. (\ref{eq:seaqiso}), which depends on the energy and entropy-energy fluctuations, $A_{ee}$ and $A_{se}$, respectively, is almost constant. The intensive property $\beta_R$ in Eq. (\ref{eq:seaqnoniso}) is identically constant. 

\begin{figure}[htbp]
   \centering
   a) \includegraphics[height=\FigSize in]{ 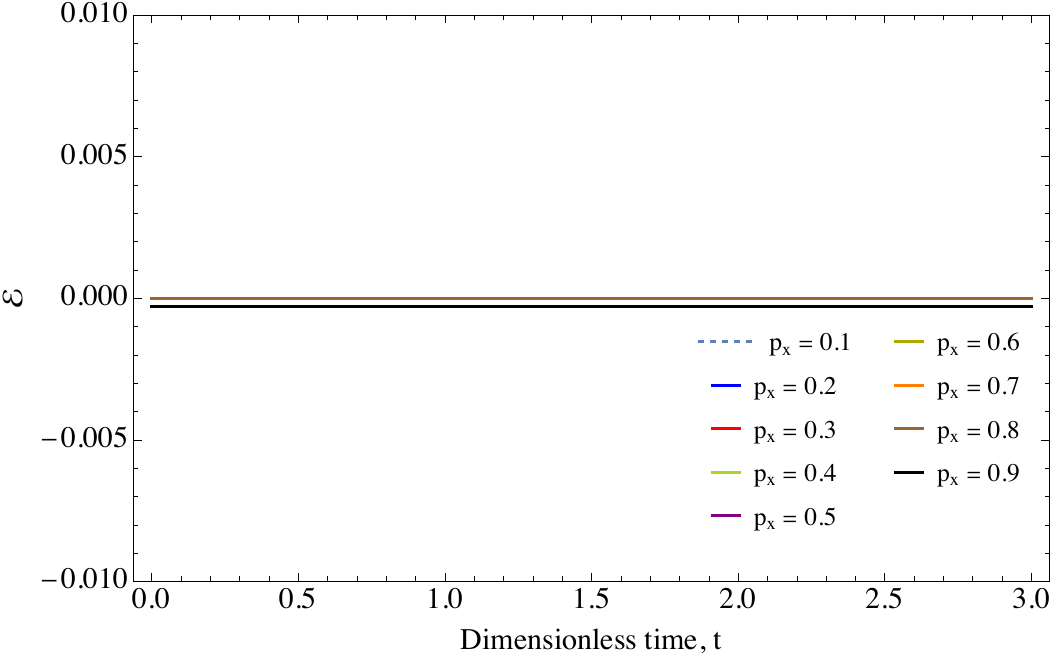} 
   b) \includegraphics[height=\FigSize in]{ 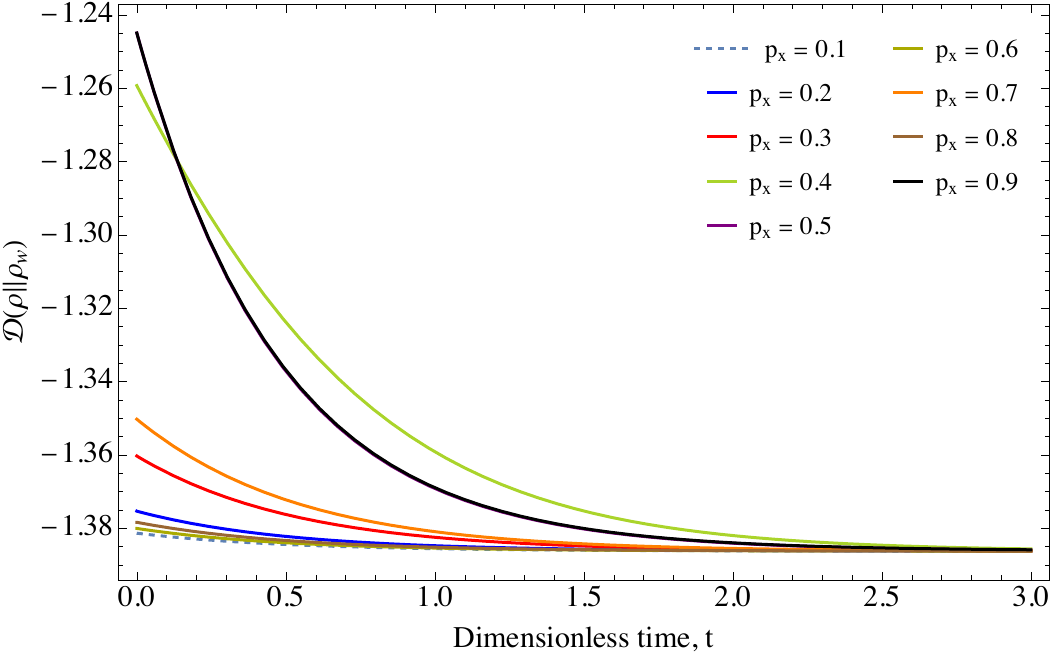} \\
   c) \includegraphics[height=\FigSize in]{ 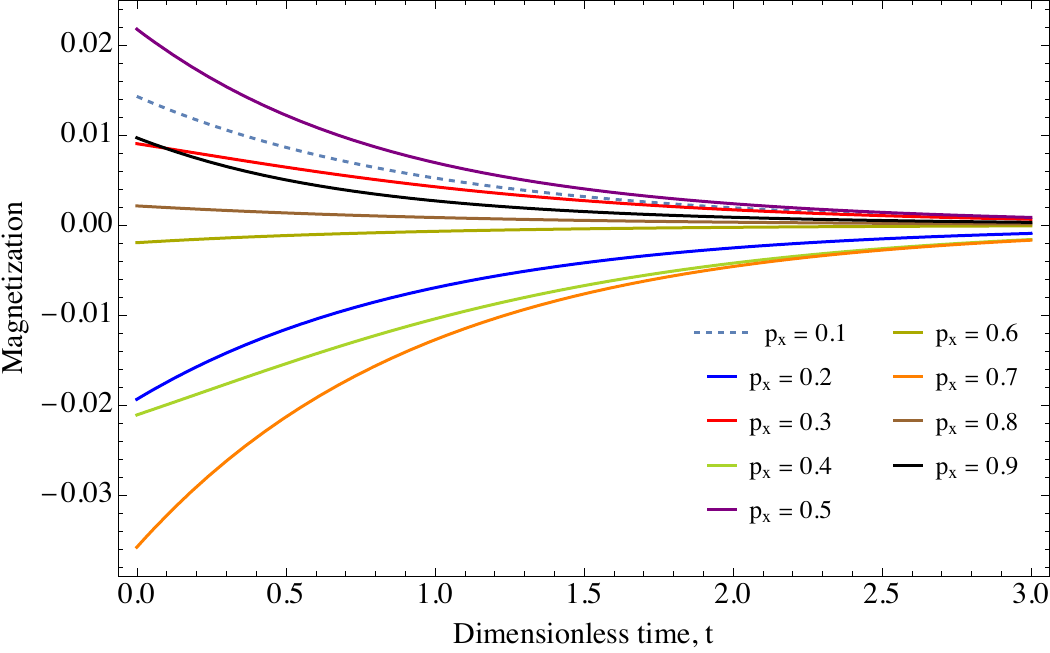} 
   d) \includegraphics[height=\FigSize in]{ 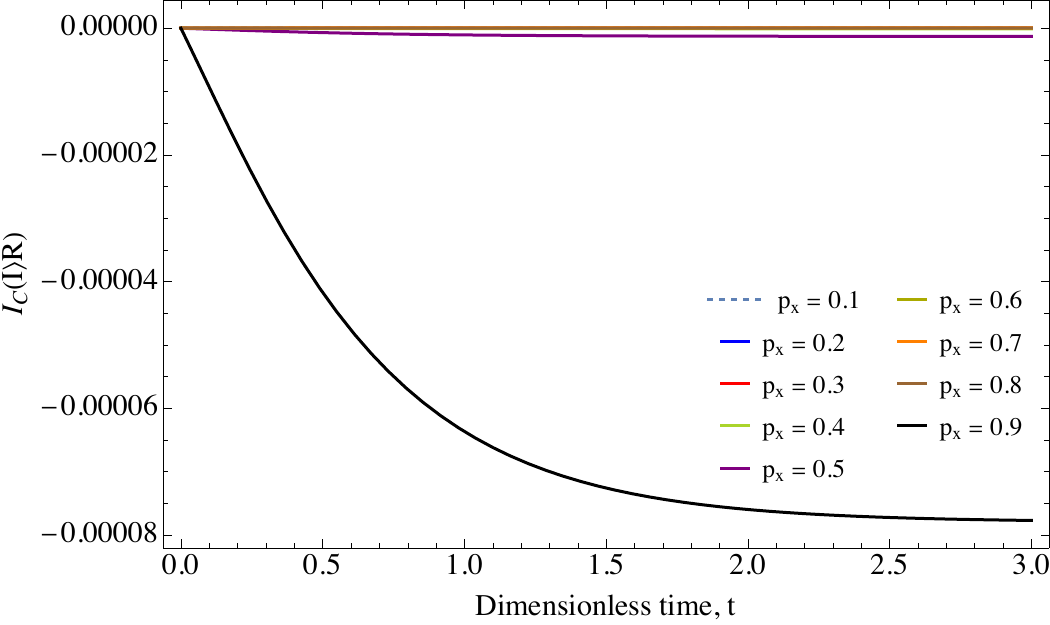} \\
   
    \caption{\label{fig:inf1} Information measure evolutions for the 2D toric code for the lattice of Fig. \ref{fig:lattice_1}: a) logarithmic negativity, b) relative entropy, c) magnetization, and d) coherent information versus time for nine perturbations. The curves for $p_x = 0.1$ begin with an initial state closest to the ground state, while those for $p_x = 0.9$ to the one furthest from the ground state for the nine perturbed initial states generated.}
  
\end{figure}

The behavior of the quantum correlations of the system are plotted in Fig. \ref{fig:inf1}a) using the logarithmic negativity. Based on the SEAQT isolated system dynamics for nine different perturbations, each initial perturbed state starts close to a zero entanglement and evolves towards a constant value of zero, leading to a sudden death of non-locality. Note that $p_x = 0.1$ in Fig. \ref{fig:inf1}a) refers to the initial state closest to the ground state of the nine initial perturbed states generated, while $p_x = 0.9$ to the one furthest from the ground state. The loss of non-locality for this case is a direct result of the dissipative dynamics of the system.

Fig. \ref{fig:inf1}b) shows the evolution of the relative entropy for the isolated system based on the SEAQT dynamics. As seen, all nine evolutions evolve to slightly different final state density operators and all are different from the ground state density operator $\hat \rho_0$. All the evolutions reach a value close to $\mathcal{D} (\hat \rho_{\text{eq}} || \hat \rho_\omega) = -1.3863$. Note that the final states are slightly different from each other because the initial perturbed states have slightly different energies from each other. Furthermore, each final state is a Gibbs (i.e., stable equilibrium) state since the SEAQT equation of motion always evolves to  stable equilibrium, i.e., for a given energy and fixed composition and parameters there is one and only one unique stable equilibrium state \cite{gyftopoulosThermodynamicsfoundationsapplications2005a}.

In Fig. \ref{fig:inf1}c), the evolution of the magnetization operator $m$ towards zero is consistent with the isolated system undergoing a relaxation process towards a Gibbs or stable equilibrium state. In this final state, the spin orientations achieve a random balance, reflecting an equilibrium condition where no particular spin orientation dominates, leading to a net magnetization of zero. All the perturbations eventually reach a magnetization value of zero once a unique stable equilibrium state is reached. 

The behavior of the coherent information seen in Fig. \ref{fig:inf1}d) involves a comparison between the entropy of the isolated system and that of the system interacting with a thermal reservoir. Based on the SEAQT model, the initial quantum information contained by the isolated system is gradually destroyed due to internal irreversibilities, while that for the system interacting with a reservoir is internally destroyed as well as transferred from the system to the reservoir. As can be seen, the coherent information approaches a constant value for each curve as the system evolves towards stable equilibrium. The constant value is different for each curve because the final stable equilibrium state for the isolated system is different for each evolution. Furthermore, a possible explanation for why the $p_x = 0.9$ evolution is significantly different from all the others is that the energy interaction with the reservoir becomes more non-linear the closer the initial perturbed state is to the original excited state, resulting, in this case, in a much larger variation in the coherent information. The dynamics for all the evolutions highlight the intrinsic tendency of quantum systems to evolve towards predictable stable equilibrium states, regardless of their initial states. These results and those in Figs. \ref{fig:inf1}a), b), and c) are consistent with the second law of thermodynamics as already recognized by Hatsopoulos and Keenan \cite{Hatsopoulos:1966} and Gyftopoulos and Beretta \cite{Gyftopoulos:2005,gyftopoulosThermodynamicsfoundationsapplications2005a}. 
\subsection{Lattice of $\text{dim} \mathcal{H} = 2^4$}
\begin{figure}[htbp]
   \centering
 a)  \includegraphics[height=1.3 in]{ 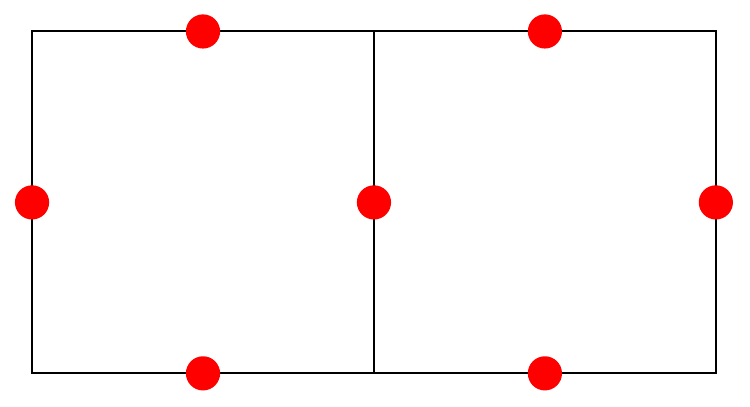}   b)  \includegraphics[height=1.3 in]{ 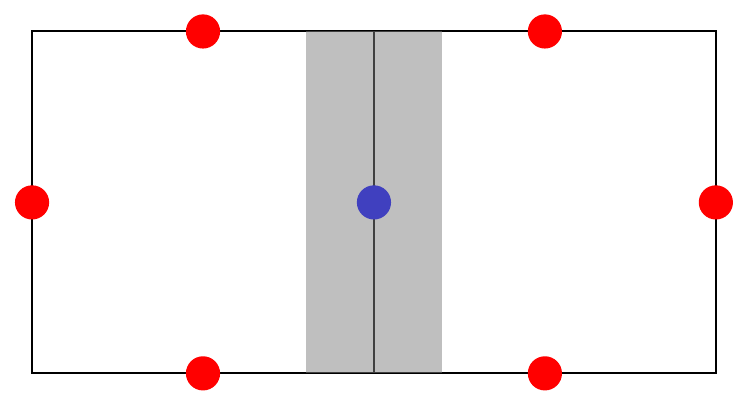} 
   \caption{\label{fig:lattice_2} Schematic of the lattice for the 2D toric code with two plaquettes in a) the ground state where the red circles represent the spin down particles in the computational basis and in b) an excited state where the blue circles represent the spin up particles in the same basis.} 
   
\end{figure}

\begin{figure}[htbp]
   \centering
   \includegraphics[scale=\FigScaleFact]{ 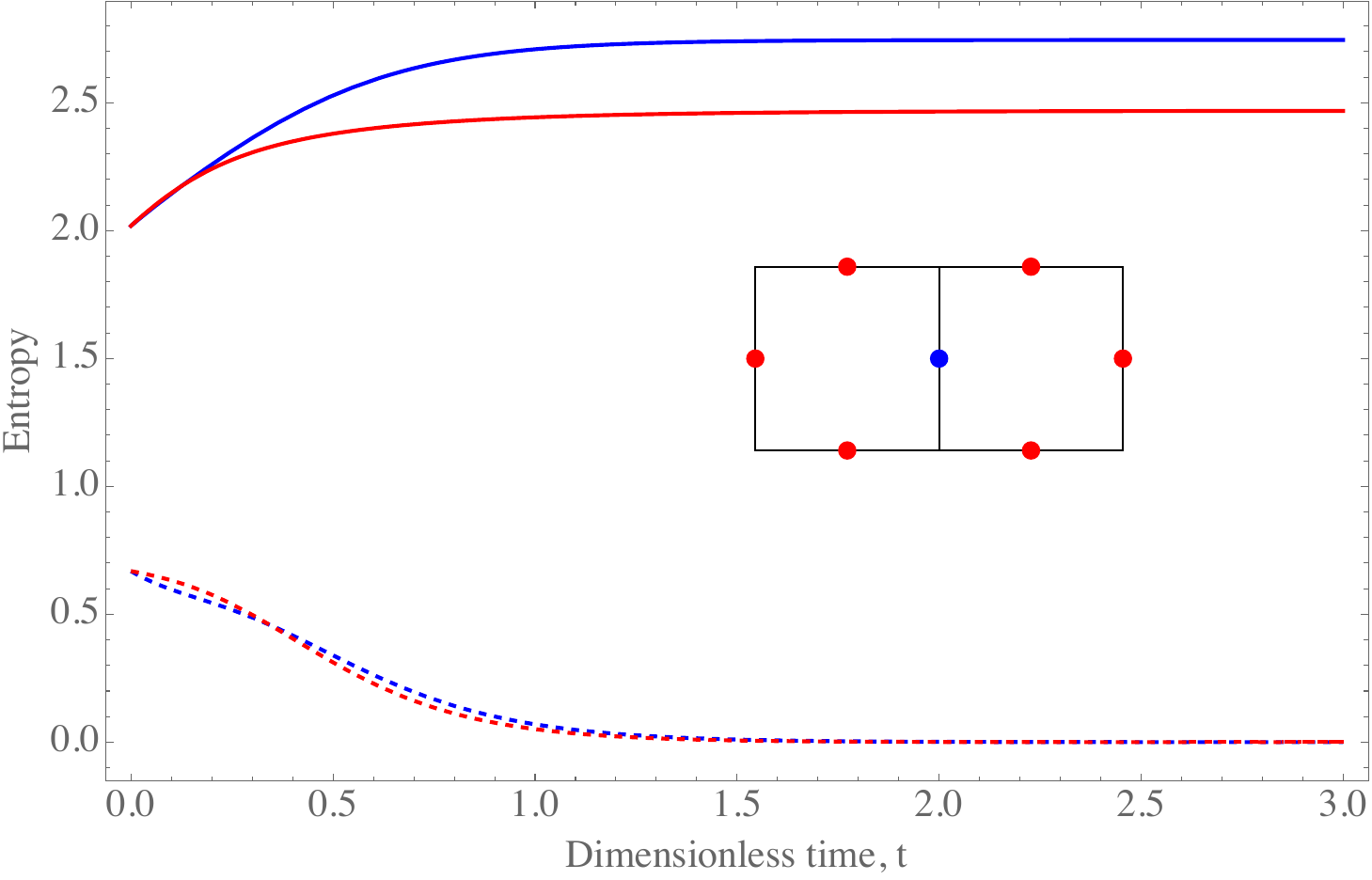} 
   \caption{ \label{fig:ent2}Entropy (solid curve) and rate of entropy change (dashed curve) of the 2D toric code with two plaquettes as a function of dimensionless time where the blue curves are for the isolated system and the red curves for the system interacting with a thermal reservoir. The initial state used is based on $p_x = 0.4$.}
  
\end{figure}

Consider now the the case of the 2D toric code with two edges represented by the lattice given in Fig. \ref{fig:lattice_2}.
Only two plaquettes and two vertex operators are considered so that the ground state density opertor takes the form
\begin{equation}
    \hat \rho_0 = \frac{1}{2}\left( \frac{\hat I+\hat A_1}{2} \frac{\hat I+\hat A_2}{2}\right)\left( \frac{\hat I+\hat B_1}{2} \frac{\hat I+\hat B_2}{2}\right)
\end{equation}
Employing the same perturbation method used previously to generate an initial state, the isolated system is allowed to evolve based on Eq. (\ref{eq:seaqiso}) as is the system interacting with a reservoir based on Eq. (\ref{eq:seaqnoniso}). The evolutions of the entropy and rate of entropy change are shown in Fig. \ref{fig:ent2} for the case of the isolated system (blue curves) and for the case of the system interacting with a thermal reservoir (red curves). The initial state used is that based on $p_x = 0.4$. In contrast to what was seen previously in Fig. \ref{fig:ent1}, the red and blue curves are distinctly different for the two evolutions since the variation in the intensive property $\beta$ in Eq. (\ref{eq:seaqiso}), which depends on the energy and entropy-energy fluctuations, $A_{ee}$ and $A_{se}$, respectively, is more significant in this case while the intensive property $\beta_R$ in Eq. (\ref{eq:seaqnoniso}) is still identically constant. Stable equilibrium is reached for both evolutions during the first 3 dimensionless time units. 

\begin{figure}[htbp]
   \centering
   a) \includegraphics[height=\FigSize in]{ 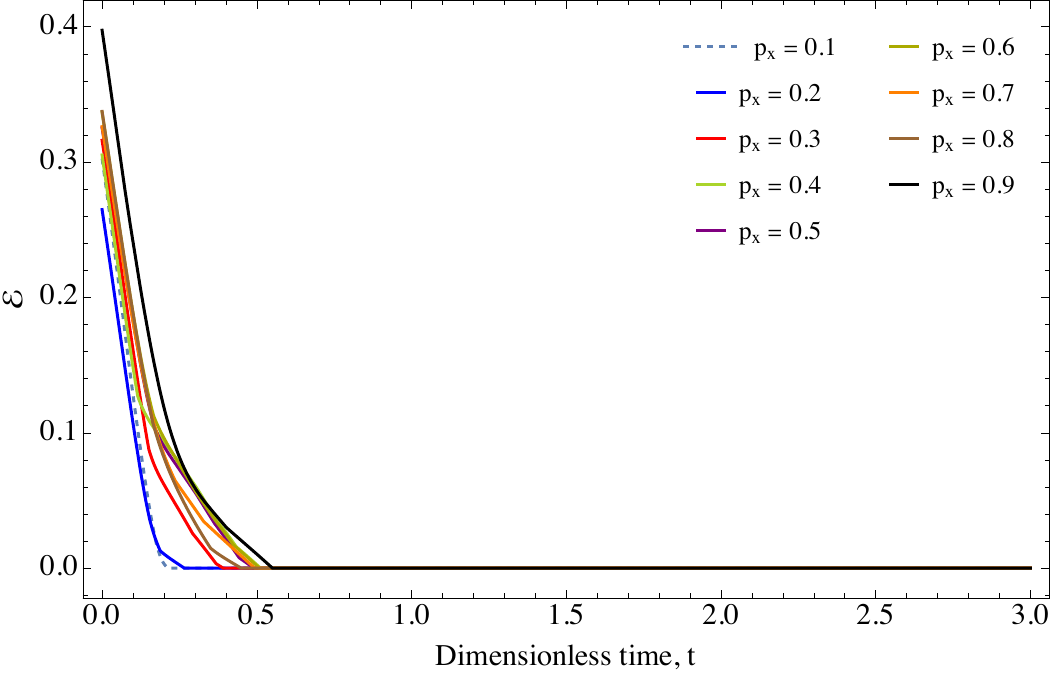} 
   b) \includegraphics[height=\FigSize in]{ 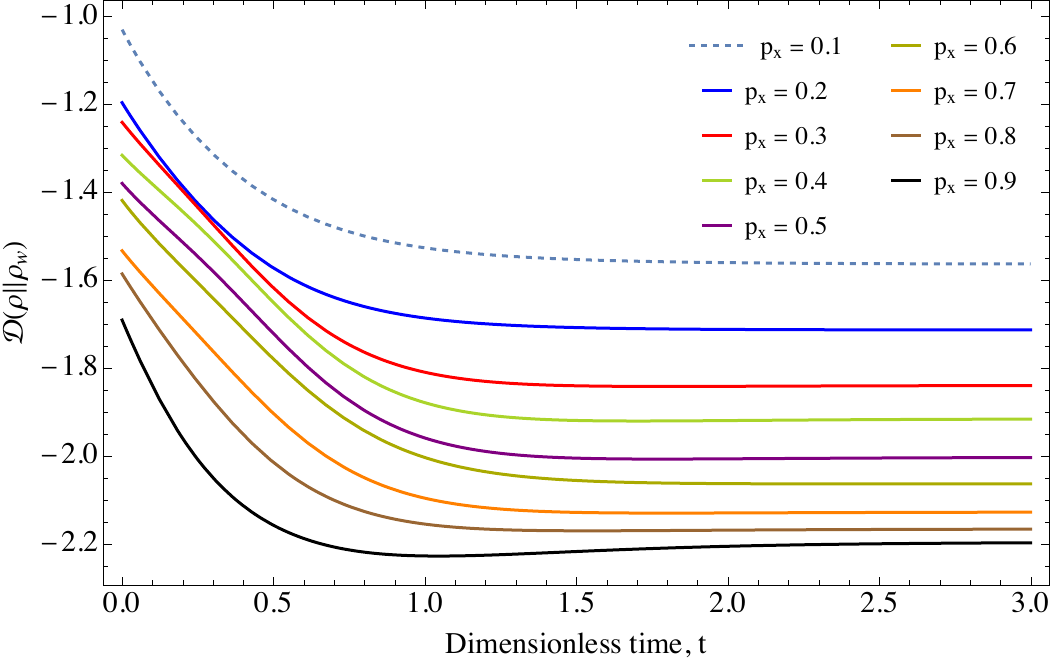} \\
   c) \includegraphics[height=\FigSize in]{ 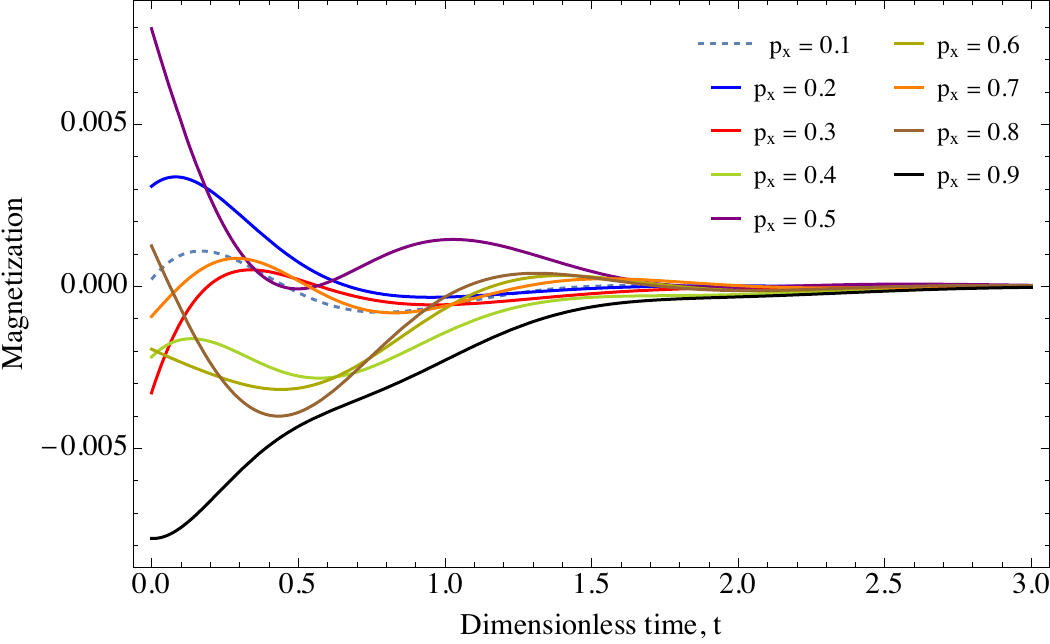} 
   d) \includegraphics[height=\FigSize in]{ 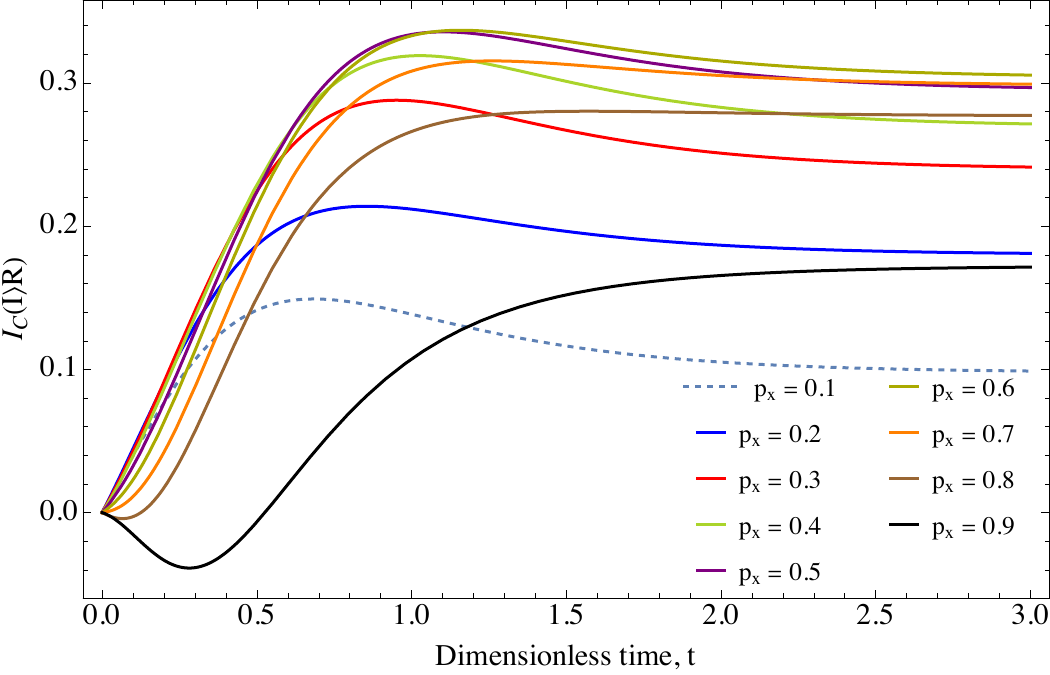} \\
   
   \caption{\label{fig:inf2}Information measures for the 2D toric code of Fig. \ref{fig:lattice_2} with two plaquettes: a) logarithmic negativity, b) relative entropy, c) magnetization, and d) coherent information versus time for nine perturbations. The curves for $p_x = 0.1$ begin with an initial state closest to the ground state, while those for $p_x = 0.9$ to the one furthest from the ground state for the nine perturbed initial states generated.}
   
\end{figure}

The behavior of all of the information measures relative to the 2D toric code of Fig. \ref{fig:lattice_2} is shown in Figs. \ref{fig:inf2}a), b), and c) for the case of the isolated system and in Fig. \ref{fig:inf2}d) for the case of the isolated system and the system interacting with a reservoir. The initial states used by the SEAQT equation of motion are generated in the usual manner \cite{Montanez:2022b} 
except that this time the ground state density matrix is perturbed by the Pauli operator $\hat \sigma_4^z$, which can be interpreted as the creation of a $z-$particle in the lattice at the position shown in blue in Fig. \ref{fig:lattice_2}. As can be seen in Fig. \ref{fig:inf2}a), the logarithmic negativity goes to zero, as expected, for all initial states.
In addition, Fig. \ref{fig:inf2}b) shows that the relative entropy in general monotonically decreases to different final stable equilibrium values with the exception of the $p_x = 0.8$ and $p_x = 0.9$ evolutions for which there is an initial steep decline followed by a gradual increase to the stable equilibrium value. For the magnetization, the SEAQT framework, as seen in Fig. \ref{fig:inf2} c), predicts an oscillatory behavior, which implies that at the beginning, there exists an order in the orientation of the spins. However, this order is eventually lost to a random ordering of spins due to irreversibilities within the system, decreasing the magnetization to zero. This oscillatory behavior is related to the interaction between adjacent spins.

In the case of the coherent information, Fig. \ref{fig:inf2} d) shows a rapid increase for values of $p_x < 0.8$ followed by a gradual decrease. The decrease is due to the fact that the net effect of internal irreversibilities and the interaction with the reservoir grows relative to the effect of internal irreversibilities only, i.e., for the former, the transfer of entropy to the reservoir decreases relative to the entropy generated internally. In contrast, the evolutions for $p_x = 0.8$ and $p_x = 0.9$ initially show a decrease in the coherent information after which they rapidly and then gradually increase before plateauing. As seen, the thermalization process towards stable equilibrium for all the evolutions results in an entropy for the isolated system at each instant of time that is larger than that for the system interacting with a reservoir, i.e., for all the evolutions the coherent information remains positive. As is discussed in the next sections, this is due to the fact that for the latter system some of the entropy generated internally is transferred from the system to the reservoir via a heat interaction, which, of course, is not the case for the isolated system.

\subsection{Lattice of $\text{dim} \mathcal{H} = 2^6$}
\begin{figure}[htbp]
   \centering
 a)  \includegraphics[height=1in]{ 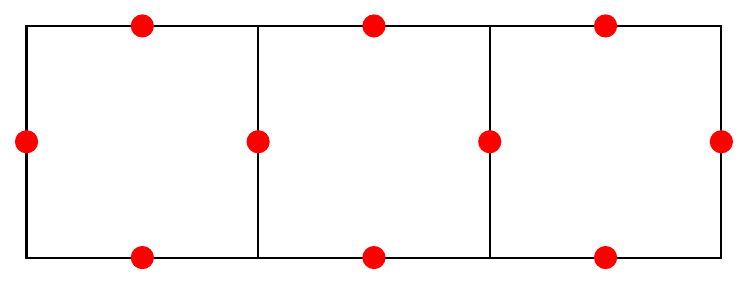}  b) \includegraphics[height=1in]{ 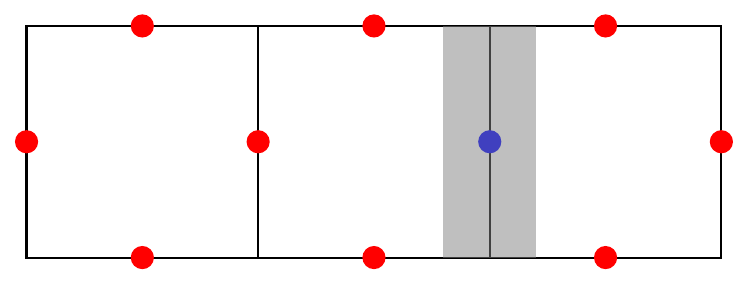}
   \caption{\label{fig:lattice_3}Schematic of the lattice of the 2D toric code with three plaquettes in a) the ground state where the red circles represent the spin down particles in the computational basis and in b) an excited state where the blue circles represent the spin up particles in the same basis.}
   
\end{figure}

\begin{figure}[htbp]
   \centering
   \includegraphics[scale=\FigScaleFact]{ 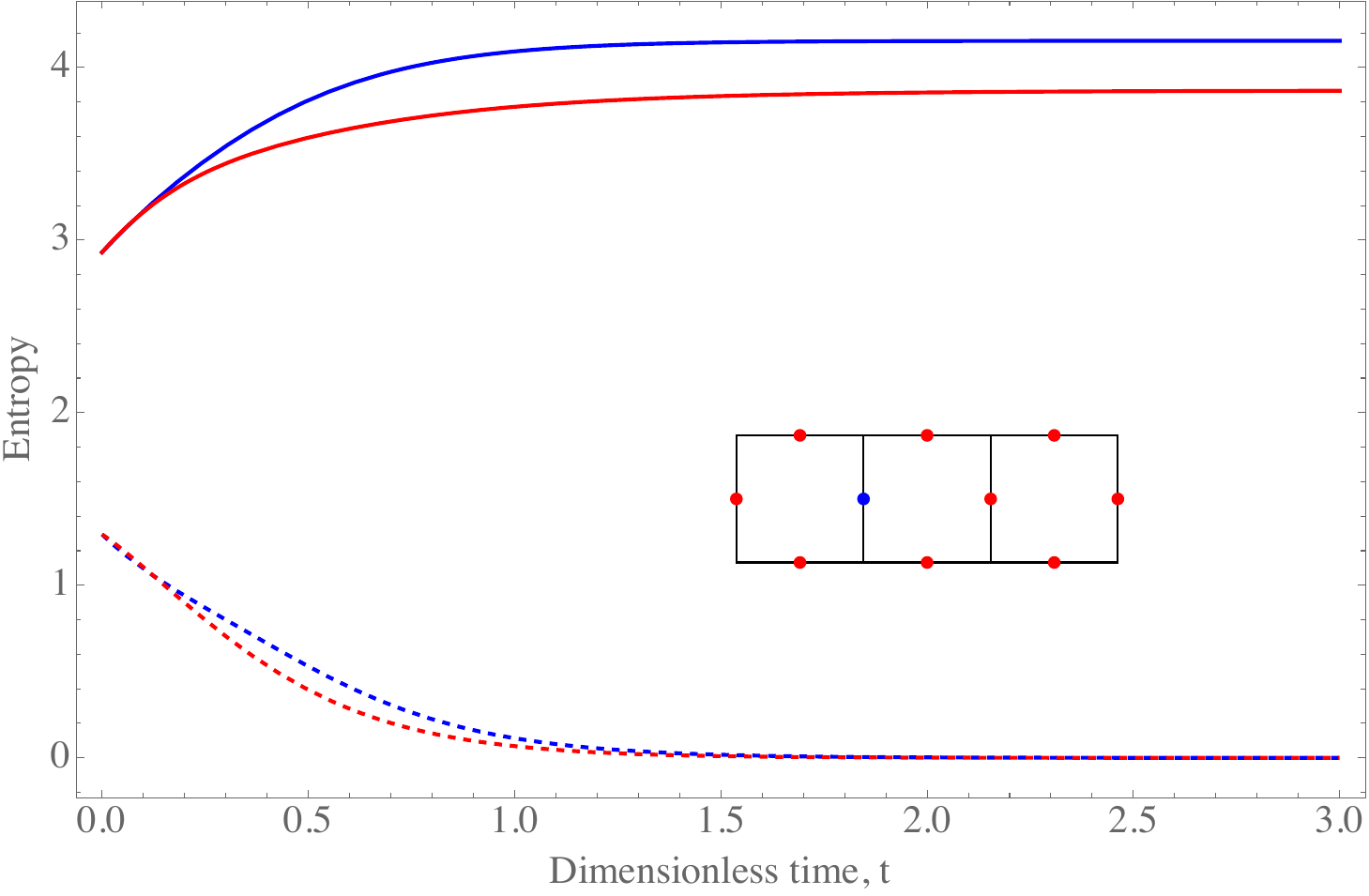} 
   \caption{  \label{fig:ent3}Entropy (solid line) and rate of entropy change (dashed line) of the 2D toric code with three plaquettes where the blue curves represent the isolated system and the red curves the system interacting with a thermal reservoir. The initial state used is based on $p_x = 0.4$.}
 
\end{figure}

 Consider now the case for two edges. The lattice for this case is shown in Fig. \ref{fig:lattice_3}, and the ground state density operator is expressed as
 \begin{equation}
    \hat \rho_0= \frac{1}{2} \left( \frac{\hat I+\hat A_1}{2} \frac{\hat I+ \hat A_2}{2}\frac{\hat I+ \hat A_3}{2} \right) \left( \frac{\hat I+ \hat B_1}{2} \frac{\hat I+ \hat B_2}{2}\frac{\hat I+ \hat B_3}{2} \right)
\end{equation}
As before, to generate the needed initial density operators, the ground state is first perturbed, but this time using a random element of a Gaussian unitary ensemble (GUE) rather than some Pauli matrix since the latter is too computationally intensive for this size lattice. The perturbation is then completed using Eq. (\ref{eq:pert}) followed by constant energy and entropy perturbations in the manner of \cite{Montanez:2022b}. 
Using the initial state generated for $p_x = 0.4$, the evolutions of the entropy as well as the rate of entropy change are predicted by the SEAQT equation of motion for the isolated system and for the system interacting with a thermal reservoir. The results are depicted in Fig. \ref{fig:ent3}. As seen, the systems reach stable equilibrium within three dimensionless time units, and the increase in the entropy for the isolated system is greater than that for the system interacting with the thermal reservoir. Again this is due to the fact that some of the entropy generated in the latter is transferred from the system to the reservoir.

\begin{figure}[htbp]
   \centering
   a) \includegraphics[height=\FigSize in]{ 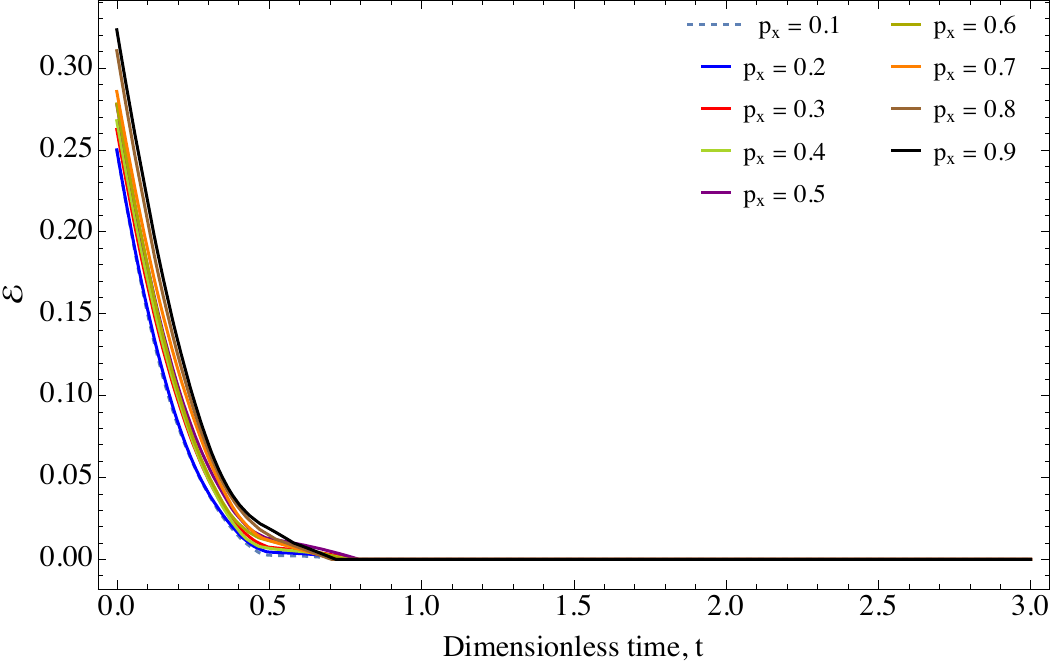} 
   b) \includegraphics[height=\FigSize in]{ 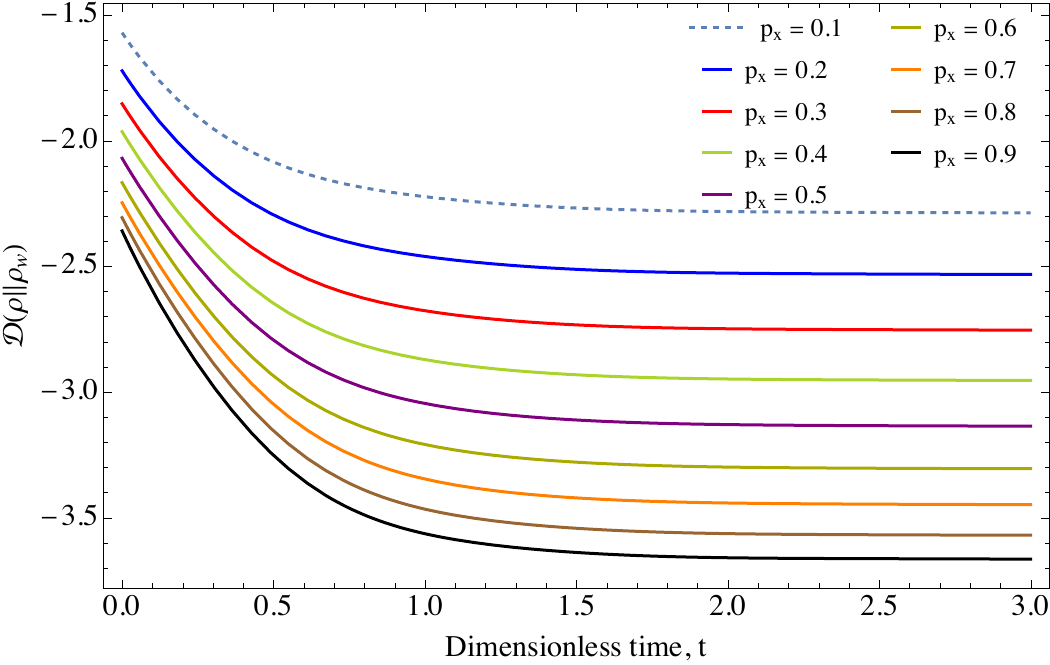} \\
   c) \includegraphics[height=\FigSize in]{ 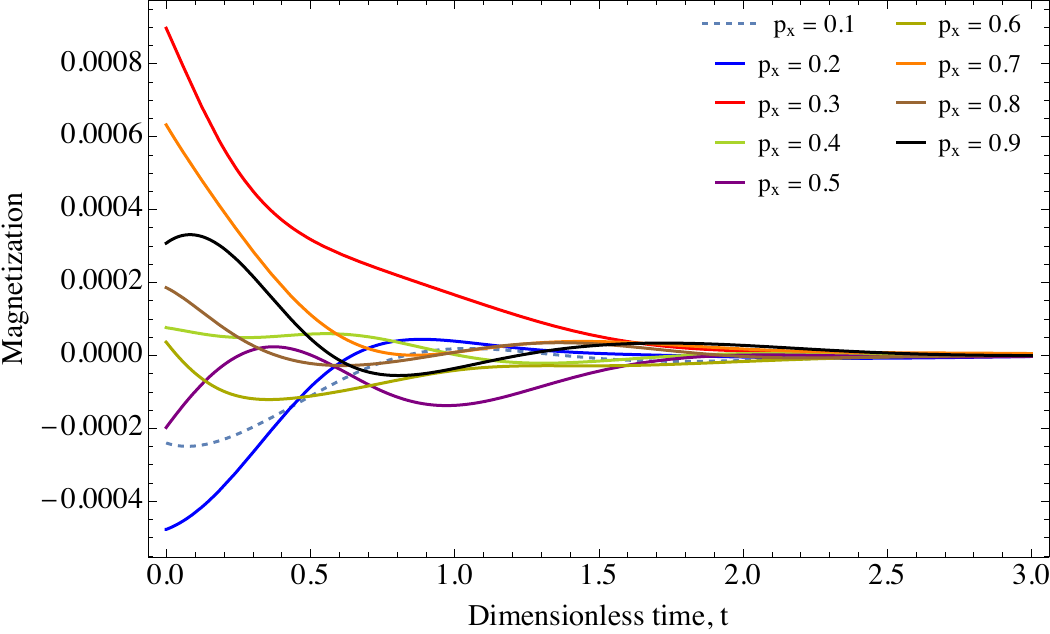} 
   d) \includegraphics[height=\FigSize in]{ 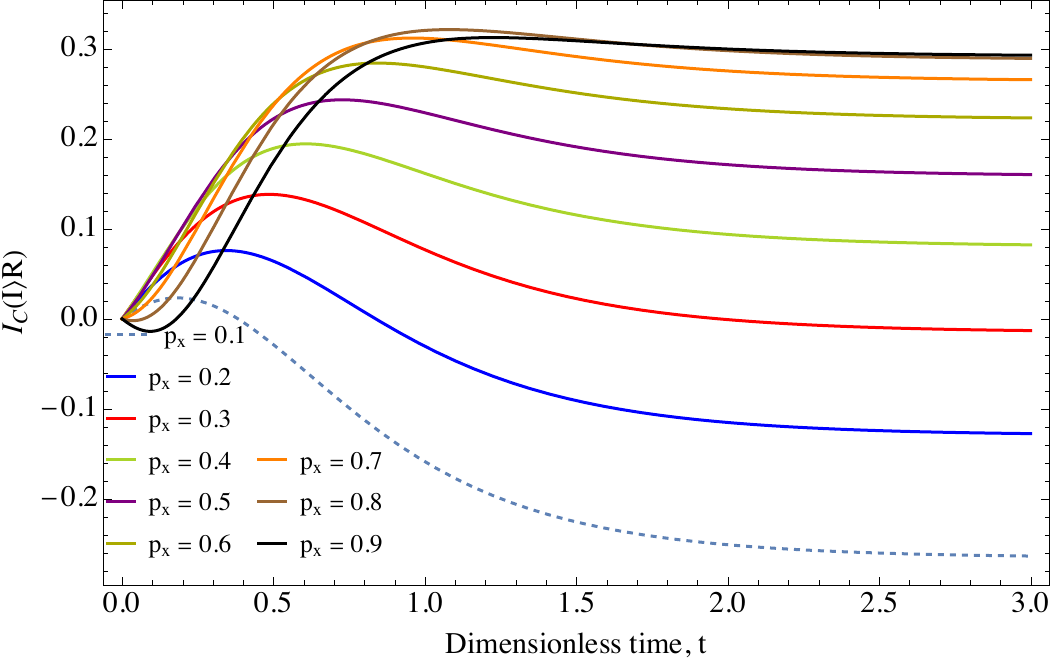} \\
   
   \caption{\label{fig:inf3} Information measures for the 2D toric code with three plaquettes: a) logarithmic negativity, b) relative entropy, c) magnetization, and d) coherent information versus time for nine perturbations. The curves for $p_x = 0.1$ begin with an initial state closest to the ground state, while those for $p_x = 0.9$ to the one furthest from the ground state for the nine perturbed initial states generated.}
   
\end{figure}

Fig. \ref{fig:inf3} depicts the dynamics towards stable equilibrium for the same four information measures utilized previously. For this analysis, the entanglement negativity with a partial transpose applied to the fourth qubit is considered. This approach is compatible with a non-contractible loop along the tori that define the lattice. As was the case in the scenario involving two plaquettes and as seen in Fig. \ref{fig:inf3}a) relative to the logarithmic negativity, the SEAQT dynamics shows a sudden death of quantum correlations for the isolated system and all nine initial states.  

As to the relative entropy, which involves a comparison between the changing excited state and the fixed ground state, Fig. \ref{fig:inf3}b) shows that, as anticipated, this measure only decreases. This reflects the increase in the entropy of the isolated system towards stable equilibrium and a density operator further and further away from that of the ground state.

Fig. \ref{fig:inf3}c) presents the evolutions of the magnetization of the 2D toric code for the isolated system and the nine different perturbations. As seen, the magnetization again oscillates around the central value of zero, approaching zero as the system approaches stable equilibrium. As noted earlier, the intrinsic irreversibilities modeled by the SEAQT equation of motion randomize the spin orientations, leading to a final state without magnetization.  

Finally, in Fig. \ref{fig:inf3}d), the coherent information shows a rapid increase for values of $p_x < 0.8$ followed by a gradual decrease. In contrast, the evolutions for $p_x = 0.8$ and $p_x = 0.9$ initially show a decrease in the coherent information after which they rapidly increase and then gradually decrease. As was seen for the previous lattice, the thermalization process for this larger lattice towards stable equilibrium immediately results in an entropy for the isolated system that is larger than that for the system interacting with a reservoir but does so only for the $p_x < 0.8$ evolutions. For the other two, the coherent information initially becomes negative and then positive.  In fact, all the evolutions for $p_x >0.2$ end up positive while the other two, i.e., those for $p_x = 0.1$ and $p_x = 0.2$ end up negative. This was not the case for the previous lattice. The negative values indicate that the entropy of the system interacting with the reservoir is larger than that of the isolated system, i.e., the net of the entropy generation and the entropy exchanged with the reservoir is larger than the entropy of the isolated system. Furthermore, at stable equilibrium, each quantum system has lost all of its initial quantum information regardless of the evolutions so that the final state of the system becomes indistinguishable from a classical statistical mixture characterized by maximal entropy.

\subsection{Lattice of $\text{dim} \mathcal{H} = 2^8$}
\begin{figure}[htbp]
   \centering
 a)   \includegraphics[height=1.5in]{ 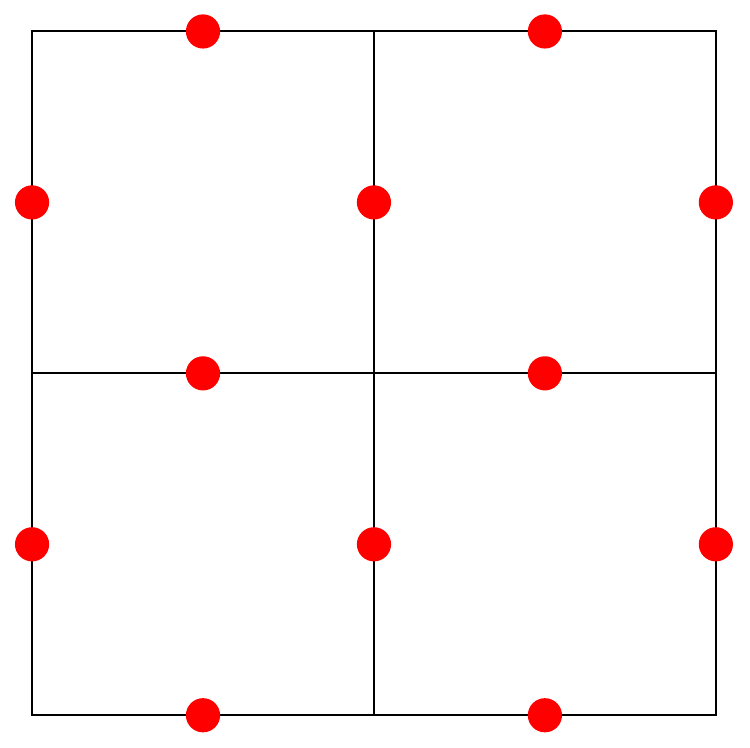}  b)  \includegraphics[height=1.5in]{ 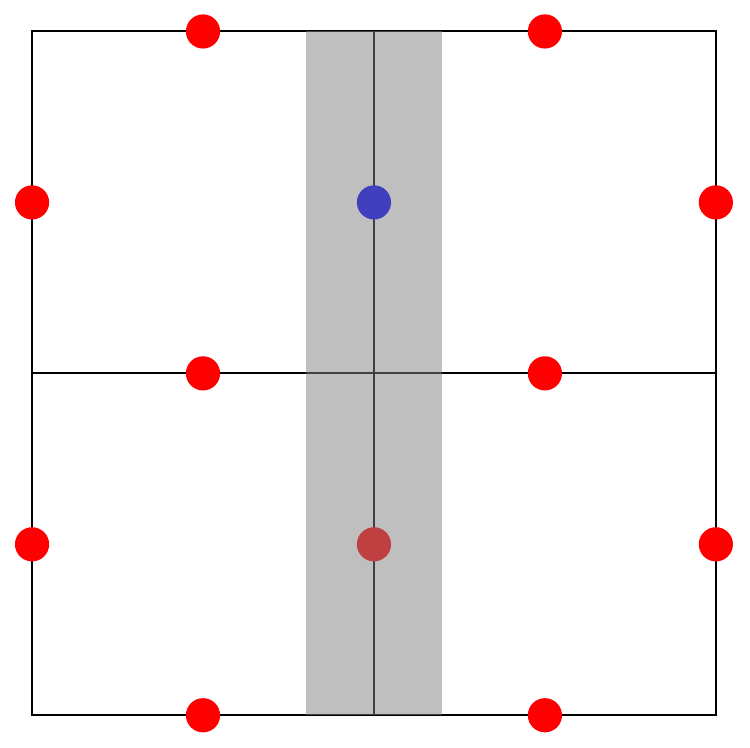}   
   \caption{\label{fig:lattice_4} Schematic of the lattice of the 2D toric code with four plaquettes in a) the ground state where the red circles represent the spin down particles in the computational basis and in b) an excited state where the blue circles represent the spin up particles in the same basis.}
   
\end{figure}

\begin{figure}[htbp]
   \centering
   \includegraphics[scale=\FigScaleFact]{ 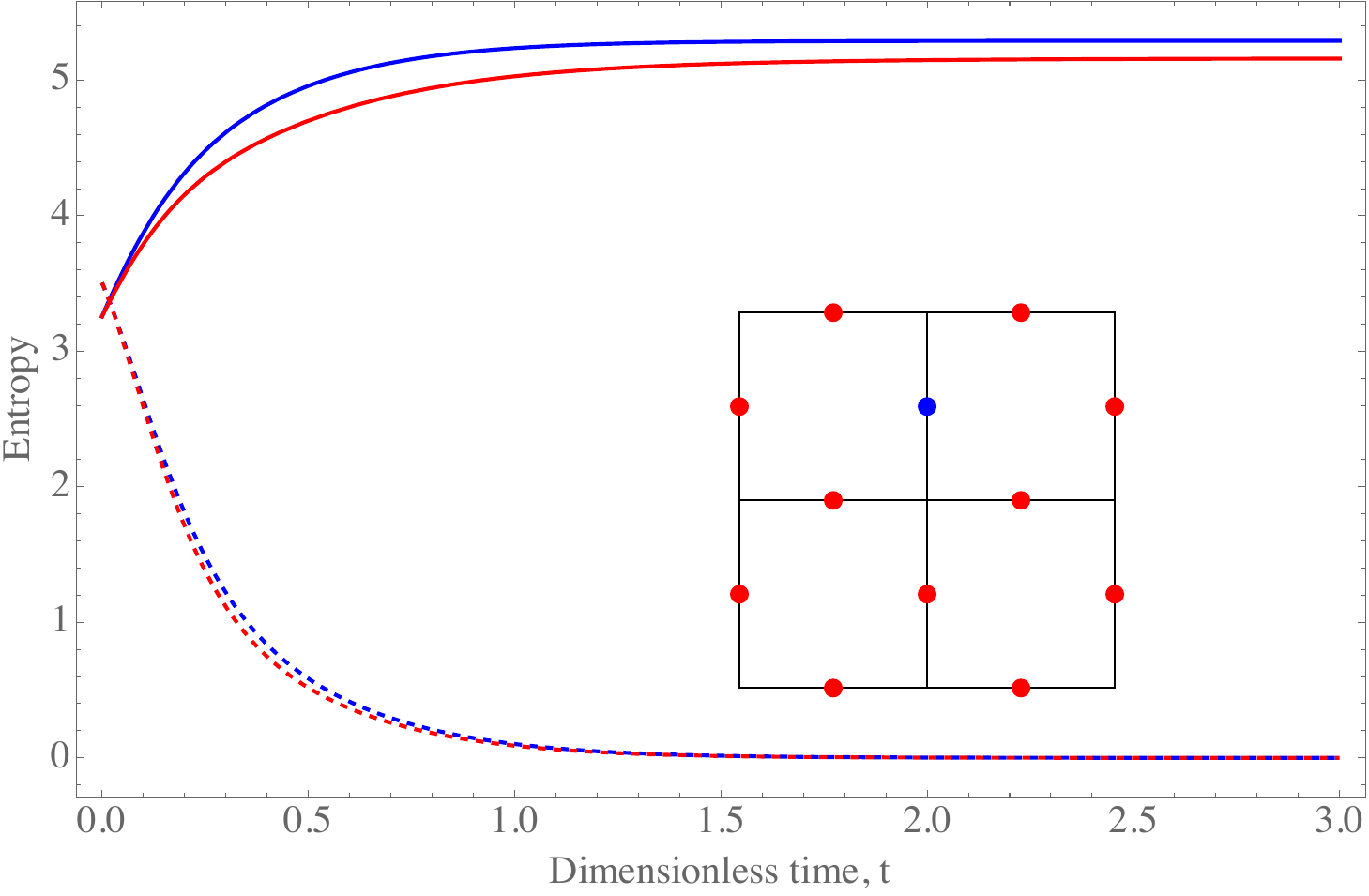} 
   \caption{ \label{fig:ent4}Entropy (solid curve) and rate of entropy (dashed curve) of the 2D toric code with 4 plaquettes where the blue curve represents the isolated system and the red curve the system interacting with a thermal reservoir. The initial state used is based on $p_x = 0.4$.}
  
\end{figure}

The final lattice considered here is the one for two edges shown in Fig. \ref{fig:lattice_4}. For this case, the ground state density operator take s the form
\begin{equation}
    \hat \rho_0 = \frac{1}{2} \left( \frac{\hat I+ \hat A_1}{2} \frac{\hat I+ \hat A_2}{2}\frac{\hat I+ \hat A_3}{2} \frac{\hat I+ \hat A_4}{2}\right) \left( \frac{\hat I+ \hat B_1}{2} \frac{\hat I+ \hat B_2}{2}\frac{\hat I+ \hat B_3}{2} \frac{\hat I+ \hat B_4}{2}\right) \,,
\end{equation}
To generate the needed initial density operators, the ground state is first perturbed, as was done previously, by using a random element of a Gaussian unitary ensemble (GUE). The perturbation is then completed using Eq. (\ref{eq:pert}) followed by constant energy and entropy perturbations in the manner of \cite{Montanez:2022b}.
Using the SEAQT equation of motion, the evolution of the entropy and the rate of entropy for the case of $p_x = 0.4$ and the isolated system is generated and the results shown in Fig. \ref{fig:ent4}. Stable equilibrium is again reached in 3 units of dimensionless time.

\begin{figure}[htbp]
   \centering
   a) \includegraphics[height=\FigSize in]{ 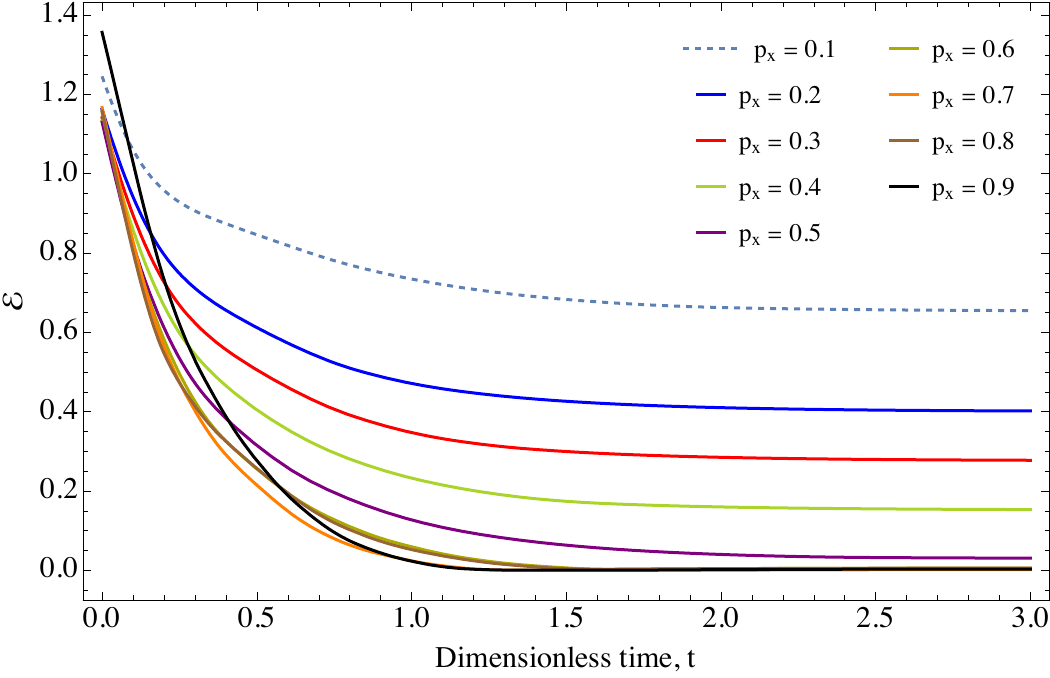} 
   b) \includegraphics[height=\FigSize in]{ 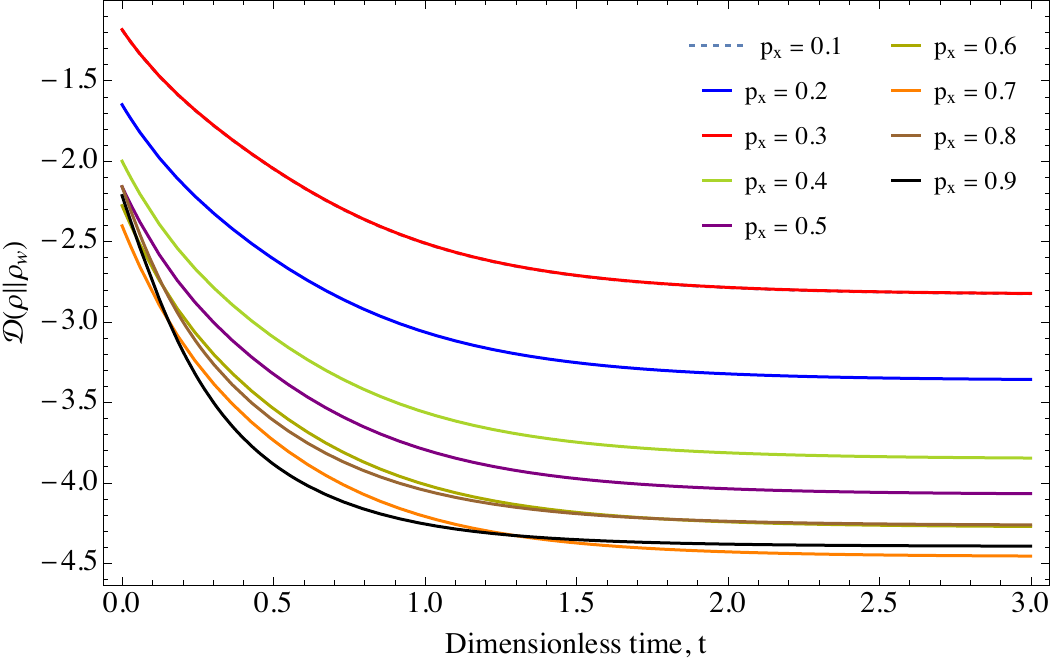} \\
   c) \includegraphics[height=\FigSize in]{ 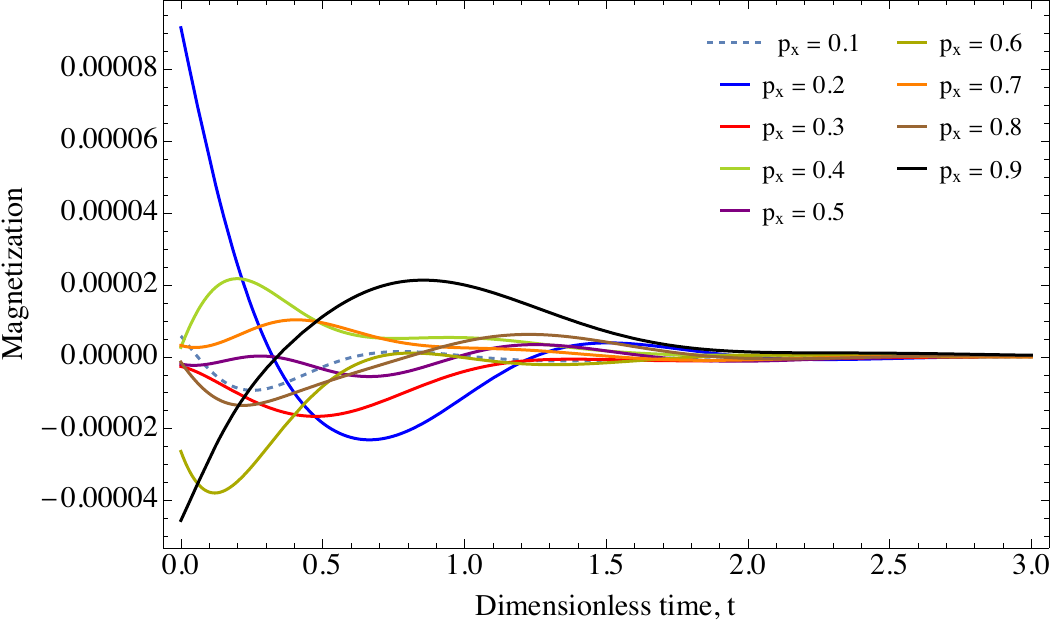} 
   d) \includegraphics[height=\FigSize in]{ 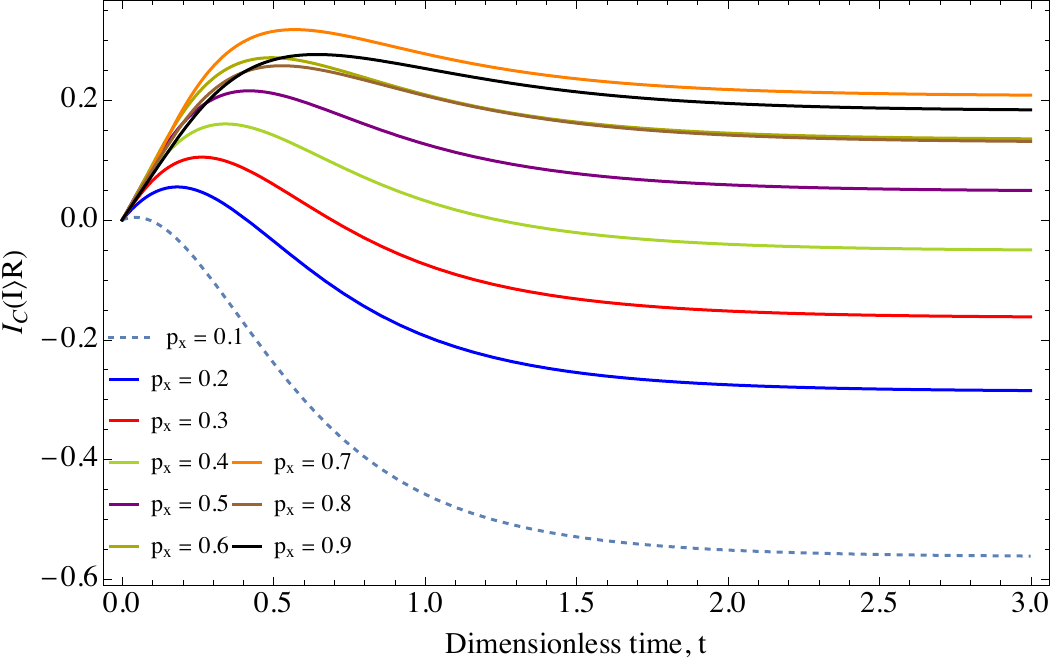} \\
   
   \caption{ \label{fig:inf4} Information measures for the 2D toric code with four plaquettes: a) logarithmic negativity, b) relative entropy, c) magnetization, and d) coherent information versus time for nine perturbations. The curves for $p_x = 0.1$ begin with an initial state closest to the ground state, while those for $p_x = 0.9$ to the one furthest from the ground state for the nine perturbed initial states generated.}
  
\end{figure}

The evolutions of the four different information measures plotted previously are shown in Fig. \ref{fig:inf4} for this lattice.
As seen in Fig. \ref{fig:inf4}a), the logarithmic negativity for the $p_x > 0.6$ evolutions goes to zero resulting in a sudden dead of quantum correlations for the isolated system. However, in contrast to the previous lattices, the evolutions for initial states closer to the ground state (i.e., for $p_x < 0.6$) conserve the quantum correlations for this lattice as the isolated system approaches stable equilibrium.

As to the relative entropy, Fig. \ref{fig:inf4}b) shows that, as was the case for the previous lattices, this measure only decreases. This reflects the increase in the entropy of the isolated system towards stable equilibrium and a density operator further and further away from that of the ground state.

The third measure, the magnetization, seen in Fig. \ref{fig:inf4}c), oscillates towards zero as the isolated system approaches to stable equilibrium. As previously noted \cite{Montantez:2020a, Montanes:2022a}, this reflects pure dephasing due to the intrinsic irreversibilities of the isolated system. At stable equilibrium, the expected zero net magnetization signifies a state of maximum entropy, where all spin orientations are equally probable, thereby minimizing the system's free energy under the given constraints.

Finally, Fig. \ref{fig:inf4}d) shows that all the evolutions experience an initial increase in the coherent information followed by a decrease. For the $p_x < 0.5$ evolutions, this decrease results in negative values of the coherent information and a loss of coherent information, i.e., the transfer of entropy to the reservoir declines relative to the entropy generated internally so that the net effect is that the entropy of the interacting system becomes larger than that of the isolated system.  

\section{\label{sec:thermo}Thermodynamics of the 2D toric code}
\begin{figure}
    \centering
    a) \includegraphics[width=2.6 in]{ 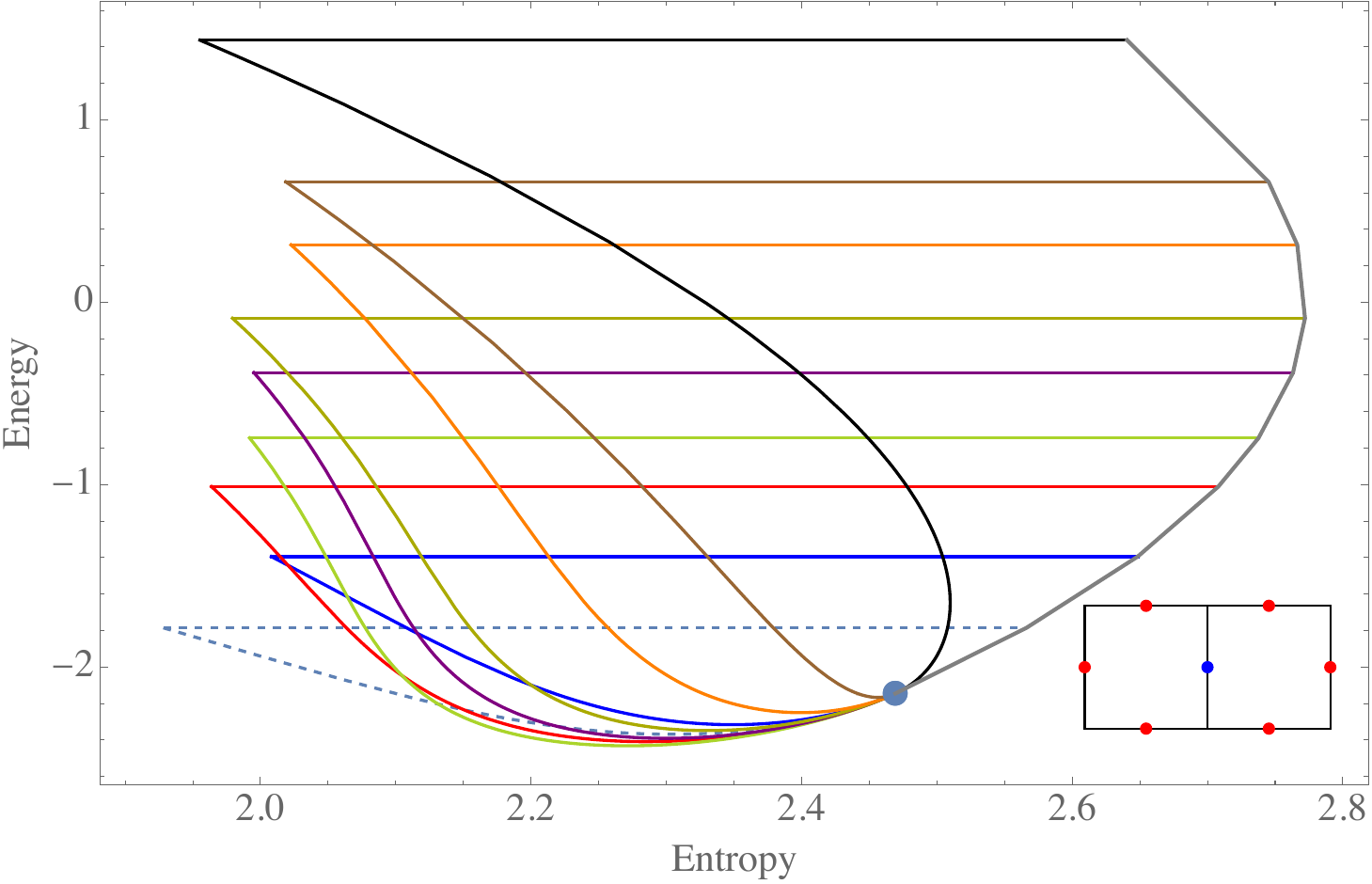}  b) \includegraphics[width=2.6 in]{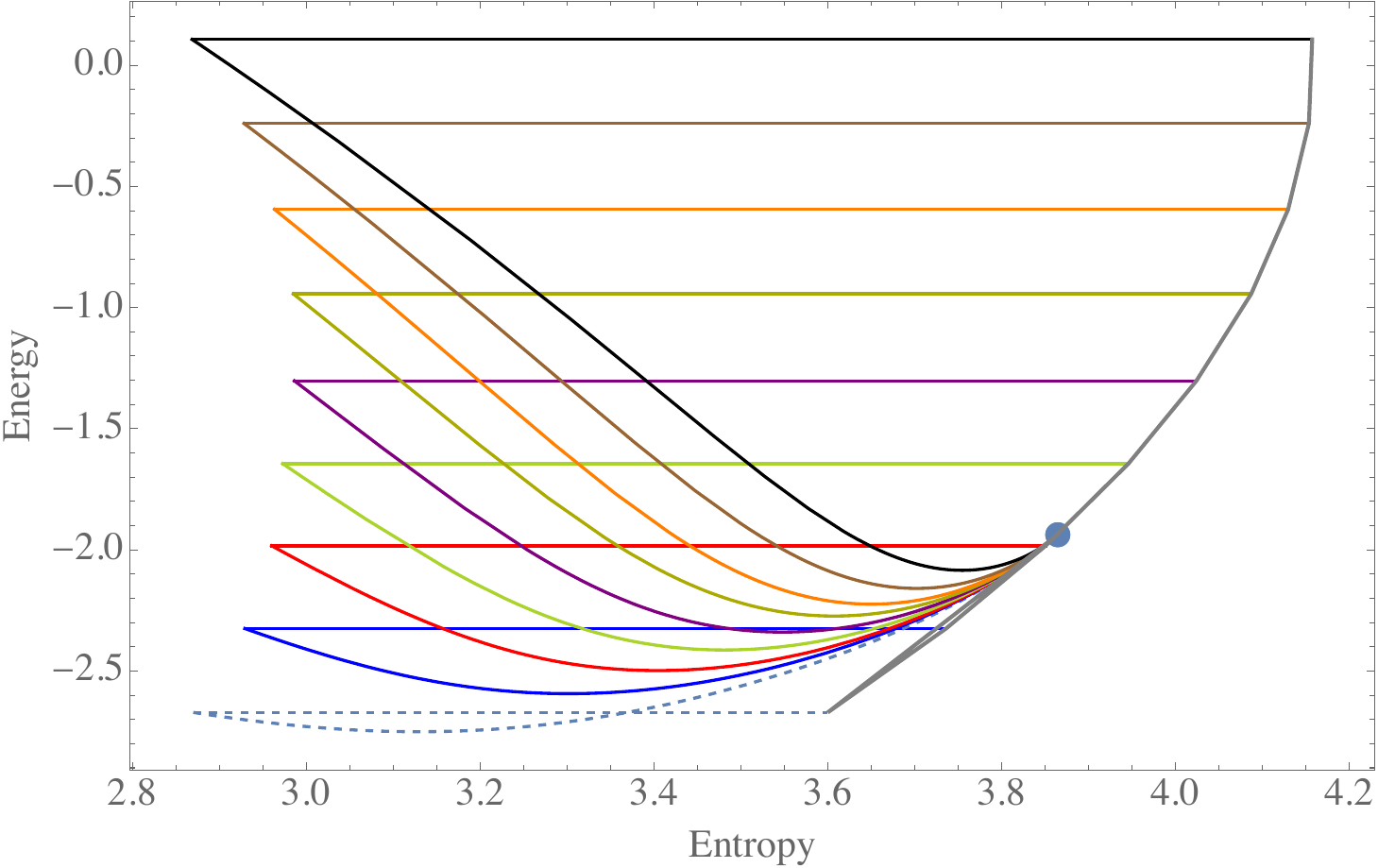}  \\
    c) \includegraphics[width=2.6 in]{ 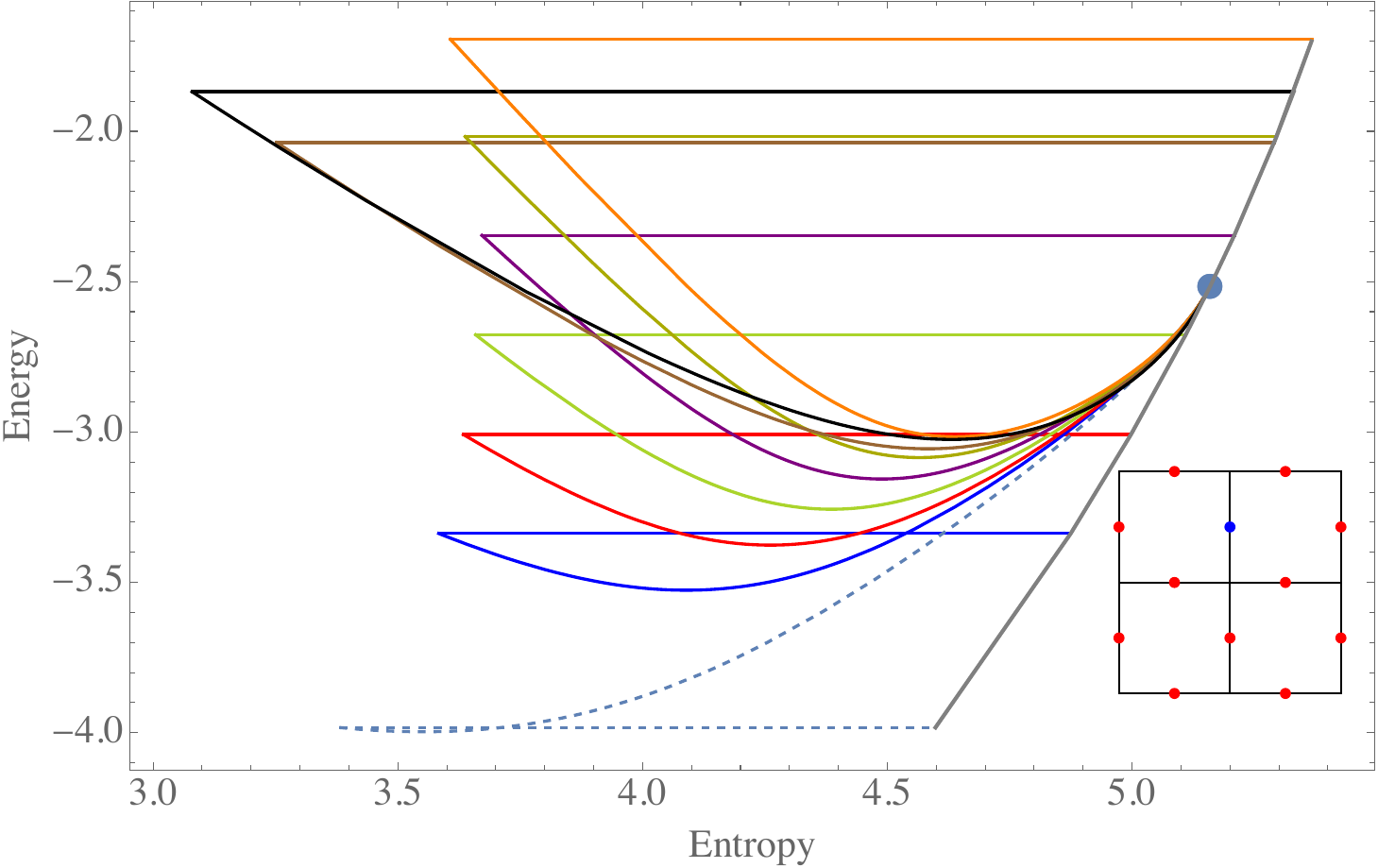}
    \caption{ \label{fig:ent-ene} Energy-entropy diagram for the lattice with a) one plaquette, b) two plaquettes, c) three plaquettes showing  the SEAQT evolutions to stable equilibrium in of the 2D toric code as an isolated system (horizontal lines) and as a system interacting with a thermal reservoir (curves converging to the blur circles). The blue circle represents the state of mutal stable equilibrium of the interacting system and the reservoir. The parabolic curve on the far right of each figure is the curve of stable equilibrium states in the energy-entropy plane.}
   
\end{figure}

In this section, the behavior of both the entropy and the geometric entropy for the three more complicated lattice configurations is examined. To begin with, Fig. \ref{fig:ent-ene} depicts, in the energy-entropy plane, the time evolution towards stable equilibrium for each configuration as an isolated system (horizontal lines) as well as a system interacting with a thermal reservoir (curves converging to the blue circle). Thus, there are two evolutions for each initial state $p_x = 
 0.1$ to $p_x = 0.9$ for a total of 18 evolutions for each lattice configuration. All the evolutions for the isolated system converge to the curve of stable equilibrium states in the energy-entropy plane depicted by the parabolic curve on the far right of Figs. \ref{fig:ent-ene}a), b), and c). Those for the interacting system converge at the blue circle, which rests on the parabolic curve. The  circle represents the final state of the interacting system when it comes to mutual stable equilibrium with the thermal reservoir. The color scheme for these curves is the same as tht used in the previous section with black corresponding to $p_x = 0.9$ and the dotted blue to $p_x = 0.1$. 

For the evolutions of the isolated system in Fig. \ref{fig:ent-ene}, the system energy remains constant and the relaxation to stable equilibrium is strictly due to an increase in the entropy caused by the entropy generation. In contrast, for the interacting system, there is a net entropy increase caused by the entropy generation and the exchange of entropy with the thermal reservoir due to a heat interaction. For the case when the energy of the system is decreasing, this involves a transfer of both energy and entropy to the reservoir, while for an increase in energy, the transfer of both energy and entropy is from the reservoir to the system. In Figure \ref{fig:ent-ene}a), the system's interaction with the environment is strictly that of the system cooling for the $p_x = 0.8$ and $p_x = 0.9$ evolutions while for the the remaining there is initial cooling followed by heating. The changes in entropy conform to the following balance of entropy on the system:
\begin{equation}
\frac{dS}{dt} =  \beta \dot Q + \dot S_{\text{gen}}
\end{equation}
where $\dot Q$ represents the heat transferred by or to the system and $\dot{\text{S}}_{\text{gen}} > 0$ denotes the entropy generated. As illustrated in Figure \ref{fig:ent-ene}a), during the cooling phases, the heat transfer and, thus, entropy transfer to the thermal reservoir is always smaller than the entropy generated for all the evolutions except the $p_x = 0.9$ evolution since the net effect is an increase in the entropy of the system. For the latter, towards the end of its evolution, the net effect of the entropy transfer and the entropy generation is for the system to experience a decrease in entropy. In the case of heating, which is seen with the $p_x < 0.8$ evolutions, both the heat interaction and the entropy generation contribute to the increase of entropy of the system. For the evolution of states of the the two largest lattices shown in Figure \ref{fig:ent-ene} b) and c), the two systems interacting with the thermal reservoir experience both cooling and heating for all the evolutions as the entropy of both systems increases.

As to the geometric entropy, it is calculated by considering a bipartite system composed of a subsystem $A$ with a single spin and a subsystem $B$ with the remaining spins. Subsystem $A$ is chosen as the region with the single spin represented by the blue circle in each of the three lattices. 
\begin{figure}
    \centering
    a) \includegraphics[width=2.6 in]{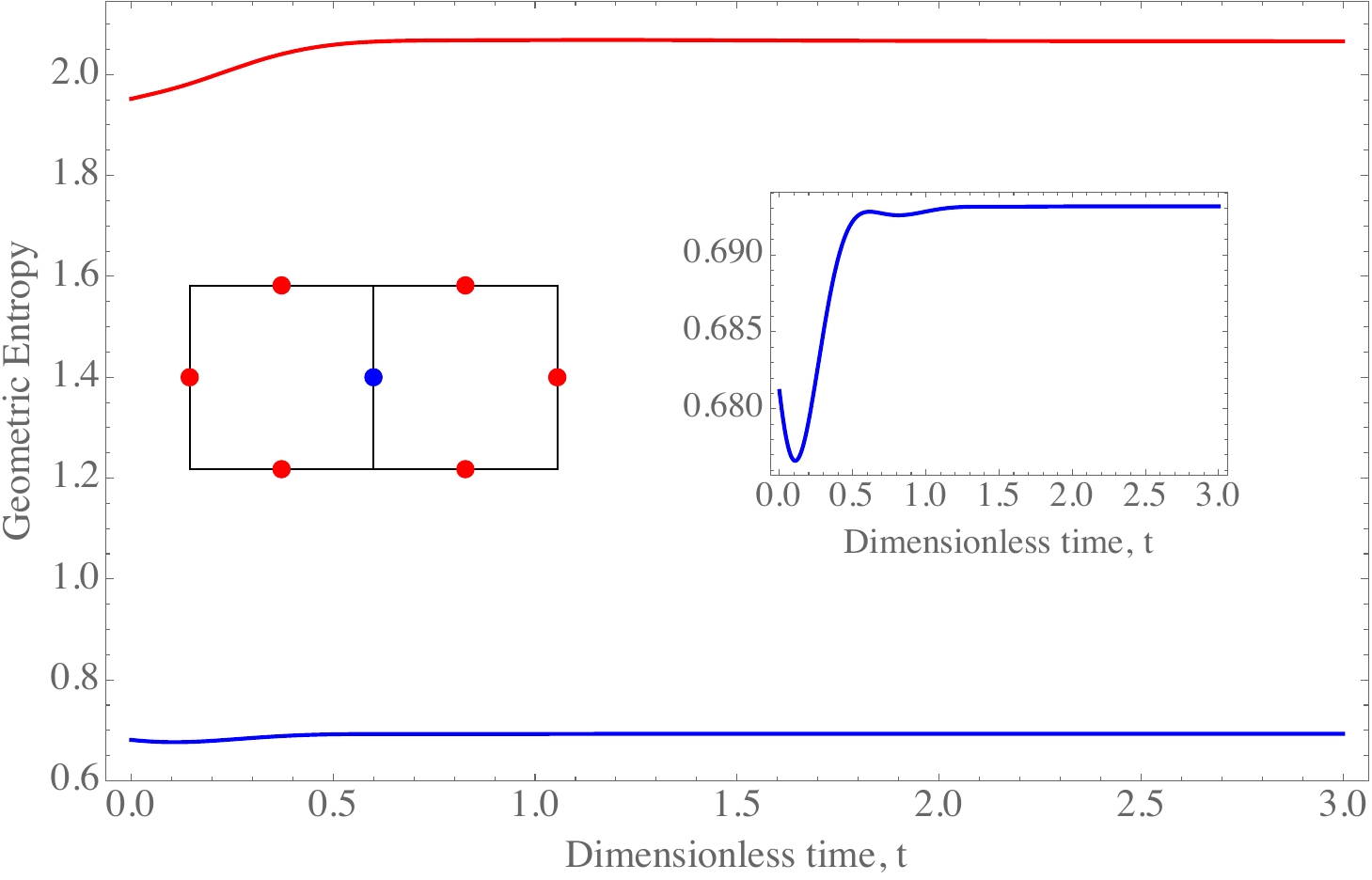}  b) \includegraphics[width=2.6 in]{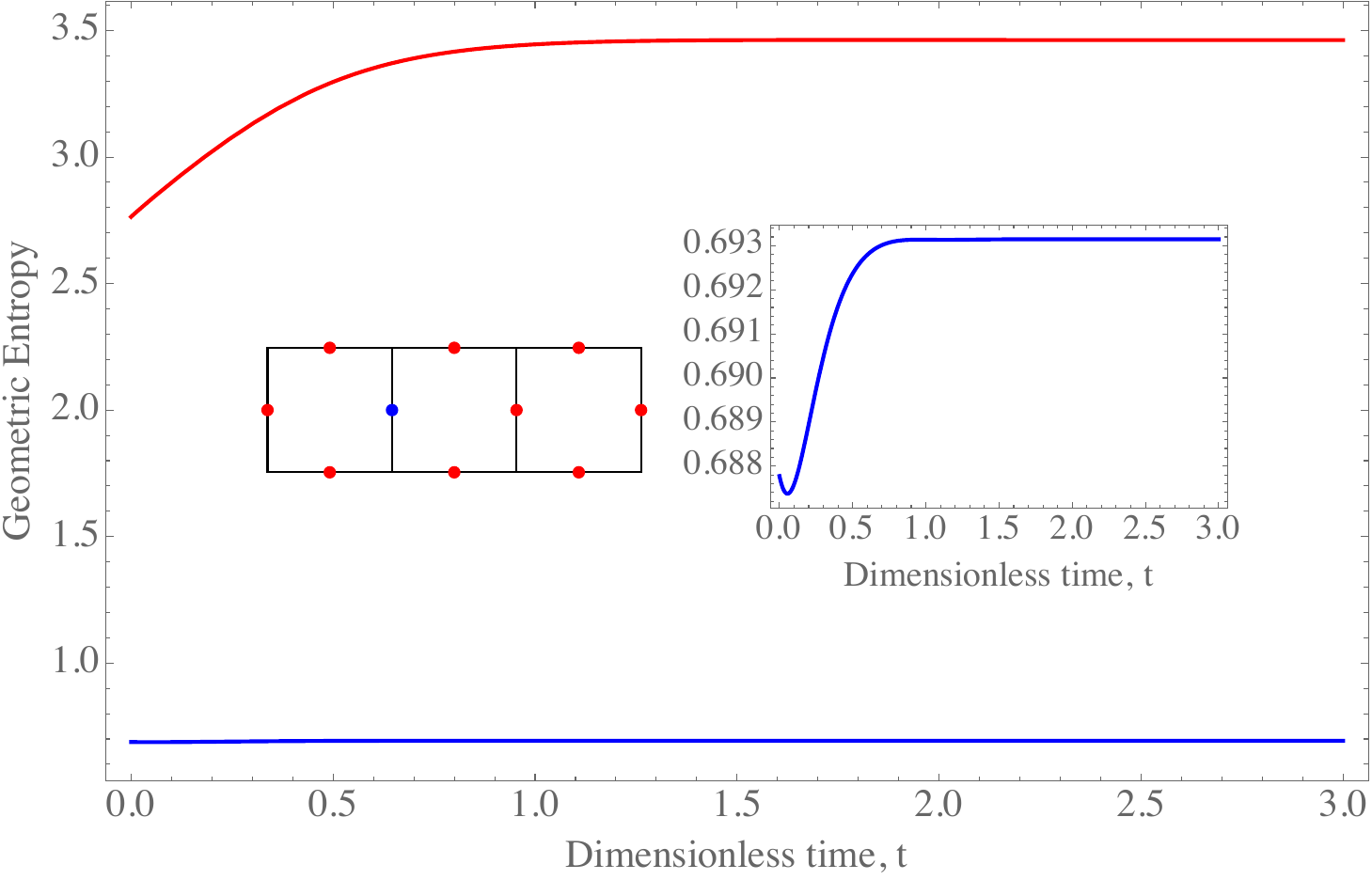}  \\
    c) \includegraphics[width=2.6 in]{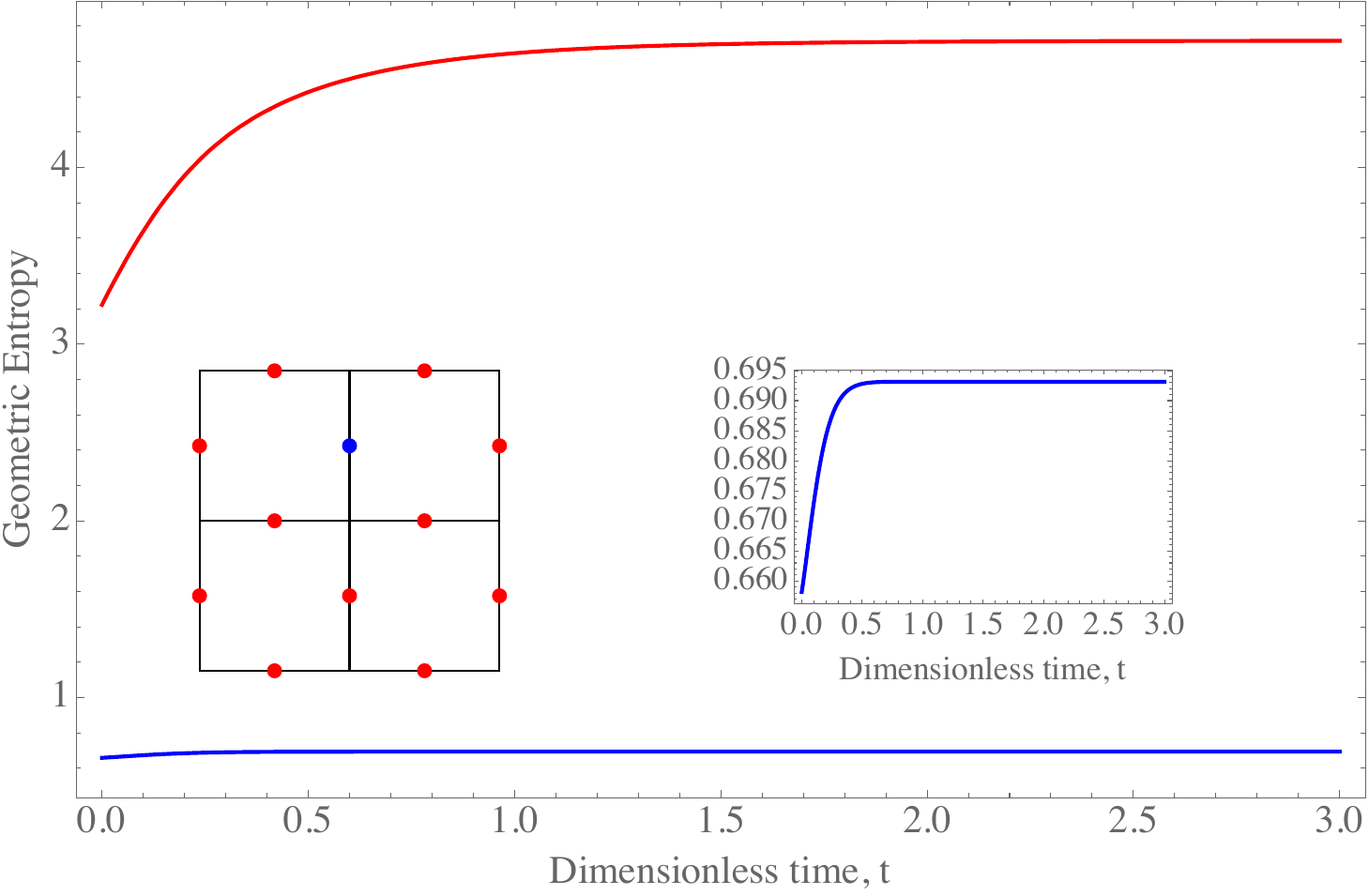}
    \caption{\label{fig:geom_ent} Geometric entropy as a function of the dimensionless time for the a) lattice with two plaquettes, b) the lattice with three plaquettes, and c) the lattice with four plaquettes where the blue curve is the geometric entropy of subsystem $A$ represented by the region which includes the blue point whereas the red curve is the geometric entropy of the region that does not include the blue point.}
    
\end{figure}
As is known \cite{Hamma:2005gr}, the geometric entropy defined as the von Neumann entropy of the reduced density matrix $\rho_A$ measures the entanglement between the degrees of freedom inside subsystem A and those outside of it.  As seen in 
Figure \ref{fig:geom_ent}a) and b), for the lattices with two and three plaquettes, there is an initial decrease in the geometric entropy after which it increases and then plateaus. Since the geometric entropy takes into account both classical and quantum correlations, this initial decrease is consistent with the fact that the quantum correlations as indicated by the logarithmic negativity decrease to zero. This death of quantum correlations is reflected in the initial decrease in the geometric entropy. Such a decrease is not seen in the largest lattice with four plaquettes because the increase in classical correlations offsets the decrease in quantum correlations. Furthermore, there is no decrease seen in the geometric entropy for subsystem $B$ for all three lattices since the effect of decreasing quantum correlations is diluted by the fact that the Hilbert space for this subsystem is much larger than that for susbsystem $A$.

Finally, it is interesting to note that at stable equilibrium, the  geometric entropy reaches a value of 
\begin{equation}
    \langle s \rangle_A =  -\text{tr}\, ( \rho_A \ln \rho_A ) = \ln 2\,,
\end{equation}
which is the expected value for a subsystem with a single spin. Thus, the SEAQT dynamics out of equilibrium produces the well-known value of the geometric entropy at stable equilibrium. 

\section{\label{sec:concl}Conclusions}
The behavior of physical systems, as described in the context of the SEAQT framework, inherently includes errors such as bit flips and pure dephasing. The 2D toric code, which is a type of quantum error-correcting code, is designed to offer fault tolerance against such errors. It can detect and correct single qubit errors efficiently, showcasing its robustness against certain types of errors. However, the  ability of the system to correct multiple, simultaneous bit flips is limited, highlighting a threshold for error correction beyond which information is irretrievably lost.

Pure dephasing errors, which lead to $\sigma_Z$ errors, are directly linked to the entropy generation in the system's dynamics. While the 2D toric code can address single qubit errors, the presence of irreversible processes degrades the system's ability to maintain coherence. This degradation points to the importance of understanding and managing the entropy in quantum systems to preserve the quantum information.  

The observation that the fluctuation of free energy is zero at stabe equilibrium underscores the intricate relationship between thermodynamic variables in quantum systems. This outcome can be attributed to the independence and lack of correlation between certain variables or to the specific distribution characteristics of these variables, as illustrated through the uniform distribution of $\langle f \rangle$ and the behavior of $\langle s \rangle^{2n}$. This insight into thermodynamic variables enhances our understanding of both the non-equilibrium and stable equilibrium properties of quantum systems and the distribution of quantum states.

Finally, SEAQT is a comprehensive framework for describing the behavior of these types of quantum system both at non-equilibrium and stable equilibrium. As shown, the SEAQT predicted stable equilibrium result for the the geometrical entropy matches the expected value of $\ln 2$, independently of the size of the lattice. However, as also shown, the SEAQT framework comprehensively predicts the behavior of these systems out of equilibrium and does so consistent with both the second and first laws of thermodynamics and the postulates of quantum mechanics, thus, adding an additional dimension to our understanding of these systems.

\section*{Acknowledgements}

A. S.-R. and C. D.-A. thank to VT for their kind hospitality during the realization of this project. A. S.-R. and C. D.-A. also thank CONAHCYT for its financial support through scholarships 252825 and 274533. 

\bibliographystyle{JHEP}

\providecommand{\href}[2]{#2}\begingroup\raggedright\endgroup

\end{document}